\documentclass{wiley-article}
\usepackage[numbers]{natbib}
\usepackage{siunitx}
\usepackage{natbib}
\usepackage{float}
\usepackage{placeins}
\usepackage{multirow}
\setcitestyle{authoryear,open={(},close={)}}

\title{Seasonal Count Time Series}

\author[1]{Jiajie Kong}
\author[2]{Robert Lund}

\contrib[\authfn{1}]{Equally contributing authors.}

\affil[1]{Department of Statistics, University of California, Santa Cruz, Santa Cruz, CA, 95064, United States}
\affil[2]{Department of Statistics, University of California, Santa Cruz, Santa Cruz, CA, 95064, United States}

\corraddress{Jiajie Kong, Department of Statistics, University of California, Santa Cruz, Santa Cruz, CA, 95064, United States}
\corremail{jkong7@ucsc.edu}

\presentadd[\authfn{2}]{Department of Statistics, University of California, Santa Cruz, Santa Cruz, CA, 95064, United States}

\fundinginfo{The authors thank NSF support from Grant DMS 21-13592.}

\runningauthor{Kong}

\begin{document}

\begin{frontmatter}
\maketitle
 
\begin{abstract}
Count time series are widely encountered in practice. As with continuous valued data, many count series have seasonal properties.  This paper uses a recent advance in stationary count time series to develop a general seasonal count time series modeling paradigm.  The model permits any marginal distribution for the series and the most flexible autocorrelations possible, including those with negative dependence.  Likelihood methods of inference can be conducted and covariates can be easily accommodated.  The paper first develops the modeling methods, which entail a discrete transformation of a Gaussian process having seasonal dynamics. Properties of this model class are then established and particle filtering likelihood methods of parameter estimation are developed.  A simulation study demonstrating the efficacy of the methods is presented and an application to the number of rainy days in successive weeks in Seattle, Washington is given.

\end{abstract}
\end{frontmatter}

\section{Introduction}

Count time series is an active area of current research, with several recent review papers and books appearing on the topic \citep{fokianos2012count, davis2016handbook, weiss2018introduction, davis2021count}. Gaussian models, which are completely characterized by the series' mean and autocovariance structure, may inadequately describe count series, especially when the counts are small.  This paper uses a recent advance in count modeling in \cite{jia2021count} to develop a very general count time series model with seasonal characteristics.  Specifically, a transformation technique is used to convert a standardized seasonal correlated Gaussian process into a seasonal count time series. The modeling paradigm allows any marginal count distribution to be produced, has very general correlation structures that can be positive or negative, and can be fitted via likelihood methods. While our most basic setup produces strictly stationary count series (in a periodic sense), nonstationary extensions, particularly those involving covariates, are easily achieved. 

With $T$ denoting the known period of the data, our objective is to model a time series $\{ X_t \}$ in time $t$ that has a count marginal distribution and periodic properties with known period $T$. A seasonal notation uses $X_{dT+\nu}$ to denote the series during the $\nu$th season of cycle $d$.  Here, $\nu \in \{ 1, 2, \ldots, T \}$ is the seasonal index and $d \in \{ 0, 1, 2, n/T-1 \}$.  We assume that there are $n$ total observations, taken at the times $1, 2, \ldots, n/T$.   To avoid trite work with edge effects, we assume that $n/T$ is a whole number.

We seek to construct count series having the cumulative distribution function $F_\nu(x)= P[X_{dT+\nu} \leq x ]$ for each cycle $d$ --- this stipulation imposes a periodic marginal distribution on the series. In fact, our constructed series will be strictly periodically stationary:  for each $k \geq 1$ and all times $t_1 < t_2 < \ldots < t_k$, the joint distribution of $(X_{t_1}, \ldots,  X_{t_k})^\prime$ coincides with that of $(X_{t_1+T}, \ldots,  X_{t_k+T})^\prime$.   We use notations such as $\{ X_t \}$ and $\{ X_{dT+\nu} \}$ interchangeably, the latter being preferred when seasonality is emphasized.

Some previous seasonal count time series models are now reviewed.  The most widely used seasonal count time series models to date develop periodic versions of discrete integer-valued autoregressions (PINAR models) --- see  \cite{monteiro2010integer, santos2019periodic}, and  \cite{bentarzi2020some}.  For example, a first order PINAR series $\{ X_t \}$ obeys the difference equation
\[ 
X_{dT+\nu} = p(\nu) \circ X_{dT+\nu-1} + \epsilon_{dT+\nu}.
\]
Here, $p(\nu) \in [0,1]$ for each season $\nu$ and $\circ$ denotes the classical thinning operator:  for an independent and identically distributed (IID) sequence of zero-one Bernoulli trials $\{ B_i \}_{i=1}^\infty$ and a count-valued random variable $C$ that is independent of $\{ B_i \}_{i=1}^\infty$, $p \circ C := \sum_{i=1}^C B_i$.  The noises $\{ \epsilon_{dT+\nu} \}$ are periodic independent and identically distributed (IID) count-valued random variables having finite second moments.  

The PINAR model class has drawbacks.  Even in the stationary case, PINAR models cannot produce some marginal distributions.  \cite{joe2016markov} quantifies the issue in the stationary case, showing that only marginal distributions in the so-called discrete self-decomposable family can be achieved.  Another issue is that PINAR models must have non-negative correlations.  Negatively correlated count series do arise \citep{kachour2009first, livsey2018multivariate, jia2021count}. Likelihood inference for PINAR and INAR models can be challenging; moreover, adding covariates to the models is non-trivial.  See \cite{joe2019likelihood} for more in the stationary setting.

A different method for constructing seasonal count series uses a periodic renewal point processes as in \cite{fralix2012renewal} and \cite{livsey2018multivariate}. Here, a zero-one binary sequence $\{ B_t \}_{t=1}^\infty$ is constructed to be periodically stationary and $\{ B_{1,t} \}, \{ B_{2,t} \}, \ldots$ denote IID copies of $\{ B_t \}_{t=1}^\infty$.  The periodic count series is constructed via the superposition
\[
X_t = \sum_{i=1}^{N_{t}} B_{i,t}.
\]
Here, $\{ N_t \}_{t=1}^\infty$ is a periodic IID sequence of count valued random variables independent of the $\{ B_{i,t} \}$.  For example, to obtain a correlated sequence $\{ X_t \}$ with Poisson marginal distributions, $\{ N_t \}$ is taken as independent Poisson in $t$, with $N_{dT+\nu}$ having the mean $\lambda_{\nu} > 0$.  Then it is easy to see that $X_{dT+\nu}$ is Poisson distributed with mean $\lambda_{\nu} P(B_{\nu}=1)$. \cite{fralix2012renewal, lund2016renewal},  and \cite{livsey2018multivariate} show how to produce the classical count marginal distributions (Poisson, binomial, and negative binomial) with this setup and consider $\{ B_t \}$ processes constructed by clipping Gaussian processes.

While binary-based models typically have negative correlations whenever $\{ B_t \}$ does, it can be difficult to achieve some marginal distributions.   A prominent example of this is the often sought generalized Poisson marginal.  Perhaps worse, likelihood methods of parameter inference appear intractable --- current parameter inference methods use Gaussian pseudo-likelihoods, which only use the series' mean and covariance structure. See \cite{davis2021count} for additional detail.

Before proceeding, a clarification needs to be made. The models constructed here posit a particular count marginal distribution for the data {\em a priori}.  This differs from dynamic linear modeling goals, where count models are often built from conditional specifications.  For a time-homogeneous AR(1) example, a dynamic linear model might employ the state space setup $X_t | \alpha_t \sim \mbox{Poisson} (e^{\alpha_t})$, where $\alpha_t = \beta \alpha_{t-1} + \eta_t$, $|\beta| < 1$, and $\{ \eta_t \}$ is zero mean Gaussian noise.  Such a setup produces a conditional Poisson distribution, not a series with a Poisson marginal distribution.  In fact, as \cite{asmussen_foss_2014} show, the marginal distribution in the above Poisson state space setup can be far from Poisson.

Additional work on periodic count series is contained in \cite{morina2011statistical, monteiro2015periodic, bentarzi2017periodic, aknouche2018periodic, santos2019theory, aknouche2018periodic}, and \cite{ouzzani2019mixture}. Most of these references take one of the above approaches.   Motivated by \cite{jia2021count}, this paper presents a different approach.

The rest of this paper proceeds as follows. The next section reviews periodic time series methods, focusing on periodic autoregressive moving-average (PARMA) and seasonal autoregressive moving-average (SARMA) difference equation structures. Section 3 clarifies our model and its properties.  Section 4 narrates parameter estimation methods and Section 5 studies these techniques via simulation. Section 6 analyzes a twenty year segment of weekly rainy day counts in Seattle, Washington.  Section 7 concludes with comments and remarks.

\section{Periodic Time Series Background}

This section briefly reviews periodic (seasonal) time series. Our future count construction uses a series $\{ Z_t \}$, standardized to $E[ Z_t ] \equiv 0$ and $\mbox{Var}(Z_t) \equiv 1$, and having Gaussian marginal distributions.  While the mean of $\{ Z_t \}$ is zero, periodic features in the autocorrelation function of $\{ Z_t \}$, which we denote by $\rho_{Z}(t,s)=\mbox{Cov}(Z_t,Z_s)$, will become paramount.

We call $\{ Z_t \}$ a PARMA($p,q$) series if it obeys the periodic ARMA($p,q$) difference equation
\begin{equation}
\label{eq:PARMA}
Z_{dT+\nu} = 
\sum_{k=1}^p \phi_k(\nu) Z_{dT+\nu-k} + 
\eta_{dT+\nu} + \sum_{k=1}^ q \theta_k(\nu) \eta_{dT+\nu-k}.
\end{equation}
Here, $\{ \eta_t \}$ is a zero mean white noise sequence with the periodic variance $\mbox{Var}(\eta_{dT+\nu})=\sigma^2(\nu)>0$.  The autoregressive order is $p$ and the moving-average order is $q$, which are taken constant in the season $\nu$ for simplicity.  The autoregressive and moving-average coefficients are $\phi_1(\nu), \ldots, \phi_p(\nu)$ and $\theta_1(\nu), \ldots, \theta_q(\nu)$, respectively, during season $\nu$.  We tacitly assume that model parameters are identifiable from the covariance of the series.  This may require more than the classical causality and invertibility conditions \citep{reinsel2003elements}.  Gaussian PARMA solutions are strictly stationary with period $T$ as long as the autoregressive polynomial does not have a root on the complex unit circle --- see \cite{lund1999modeling} for quantification.  Not all PARMA parameters are free due to the restriction $\mbox{Var}(X_t) \equiv 1$; the following example delves further into the matter.

\vspace{.12in} \noindent {\bf Example 3.1} A PAR(1) series with period $T$ obeys the recursion
\begin{equation}
\label{eq:PAR1}
Z_t = \phi(t) Z_{t-1} + \eta_t,
\end{equation}
where $\{\eta_t \}$ is zero mean white noise with $\mbox{Var}(\eta_t)= \sigma^2(t)$.  The quantities $\phi(t)$ and $\sigma^2(t)$ are assumed periodic in $t$ with period $T$.  This difference equation is known to have a unique (in mean squared) and causal solution whenever there is a stochastic contraction over an entire cycle: $| \prod_{\nu=1}^T \phi(\nu) | < 1$ \citep{lund1999modeling}.

To impose $\mbox{Var}(Z_t) \equiv 1$, take a variance on both sides of (\ref{eq:PAR1}) and set $\mbox{Var}(Z_t) = \mbox{Var}(Z_{t-1})=1$ to infer that $\sigma^2(t) = 1 - \phi^2(t)$, which we tacitly assume is positive for all $t$.  This uses $\mbox{Cov}(Z_{t-1}, \eta_t)=0$, which follows by causality.  The covariance structure of $\{ Z_t \}$ can now be extracted as 
\[
\rho_{Z}(t,s) = \prod_{i=0}^{t-s-1} \phi(t-i).
\]
for $s < t$. $\clubsuit$.

Another class of periodic models in use today are the SARMA series.  SARMA series are actually time-stationary models, but have comparatively large autocorrelations at lags that are multiples of the period $T$.  The most basic SARMA($p,q$) series $\{ Z_t \}$ obeys a difference equation driven at lags that are multiples of the period $T$:
\begin{equation}
\label{eq:SARMA}
Z_t = 
\sum_{k=1}^p \phi_k Z_{t-kT} + 
\eta_t + \sum_{k=1}^q \theta_k \eta_{t-kT},
\end{equation}
where $\{ \eta_t \}$ is zero mean independent noise with a constant variance. In this setup, $\rho_Z(t,s)=0$ unless $t-s$ is a whole multiple of the period $T$.  As such, many authors allow $\{ \eta_t \}$ to have additional correlation, specifically a zero mean ARMA($p^*,q^*$) series.  This results in a model that can have non-zero correlations at any lag; however, the model is still stationary and does not have any true periodic features.  Since the model is stationary, we write $\rho_Z(t,s)= \rho_Z(t-s)$.

\vspace{.12in} \noindent {\bf Example 3.2} A SAR(1) series with period $T$ and AR(1) $\{ \eta_t \}$ obeys the difference equation pair
\begin{equation}
\label{e:system}
Z_t = \phi Z_{t-T} + \eta_t; \quad
\eta_t = \alpha \eta_{t-1} + \epsilon_t,
\end{equation}
where $\{ \epsilon_t \}$ is zero mean white noise with variance $\sigma^2_\epsilon$, $|\phi|<1$, and $|\alpha|< 1$. Combining these two difference equations results in a stationary and causal AR($T+1$) model for $\{ Z_t \}$.   

Imposing $\mbox{Var}(Z_t) \equiv 1$ and taking a variance in the first equation in (\ref{e:system}) gives
\[
1 = \phi^2 + \mbox{Var}(\eta_t) + 
2 \phi\mbox{Cov}(Z_{t-T}, \eta_t). 
\]
To proceed, use equation (\ref{e:system})'s causal solutions $\eta_t = \sum_{k=0}^\infty \alpha^k \epsilon_{t-k}$ and $Z_{t-\ell} = \sum_{m=0}^{\infty}\phi^m \eta_{t-mT-\ell}$ to get
\begin{equation}
\label{inter1}
\mbox{Cov}( \eta_t, Z_{t-\ell} ) = \sigma^2_\epsilon \frac{\alpha^{\ell}}{(1-\alpha^2)(1-\phi\alpha^T)}
\end{equation}
for any $\ell > 0$.  Combining the last two equations, we see that taking
\begin{equation}
\label{varset}
\sigma_\epsilon^2 = \frac{(1-\phi^2)(1-\alpha^2)(1-\phi\alpha^T)}{1+\phi\alpha^T}
\end{equation}
indices $\mbox{Var}(Z_t) \equiv 1$.

To extract the covariance structure of $\{ Z_t \}$, multiply both sides of (\ref{e:system}) by $Z_{t-h}$ and take expectations to get the Yule-Walker type equations
\begin{eqnarray*}
    \rho_Z(0) &=& \phi \rho_Z(T) + E(Z_t\eta_t)\\
    &\vdots&\\
    \rho_Z(T) &=& \phi\rho_Z(0) + E(Z_{t-T}\eta_t).
\end{eqnarray*}
This system can be rewritten in vector form as
\[
\begin{bmatrix}
    1&0&\cdots&0&-\phi\\
    0&1&\cdots&-\phi&0\\
    \vdots&\vdots&\vdots&\vdots&\vdots\\
    0&-\phi&\cdots&1&0\\
    -\phi&0&\cdots&0&1
\end{bmatrix}\begin{bmatrix}
    \rho_Z(0)\\
    \rho_Z(1)\\
    \vdots\\
    \rho_Z(T-1)\\
    \rho_Z(T)
\end{bmatrix}=\begin{bmatrix}
    E(\eta_tZ_t)\\
    E(\eta_tZ_{t-1})\\
    \vdots\\
    E(\eta_tZ_{t-T+1})\\
    E(\eta_tZ_{t-T})
\end{bmatrix}.
\]
One can show that the inverse of the matrix in the above linear system exists.  From this, (\ref{inter1}), (\ref{varset}), and some very tedious algebraic manipulations, one can extract
\[
\rho_Z(h)=\frac{\alpha^h+\phi\alpha^{T-h}} 
{1 + \phi\alpha^T}, \quad 0 \leq h \leq T.
\]
For the $h>T$ model correlations, multiply the first equation in (\ref{e:system}) by $Z_{t-h}$ for $h > T$ and take expectations to get the recursion $\rho_Z(h) = \phi \rho_Z(h-T) + E(\eta_t Z_{t-h})$. This can be solved with (\ref{inter1}) to get
\[
\rho_Z(h) =  \phi^{a}\frac{\alpha^b+\phi\alpha^{T-b}}{1+\phi\alpha^T}+\sum_{k=0}^{a-1}\phi^k\alpha^{h-Tk}\frac{1-\alpha^2}{1+\phi\alpha}, \quad h>T,
\]
where $a = \lfloor h/T \rfloor$ and $b=h-aT$. $\clubsuit$.

PARMA and SARMA methods are compared in detail in \cite{lund2011choosing}. PARMA models are usually more applicable since the immediate past of the process is typically more influential than past process lags at multiples of the period $T$.  Applications in the environment \citep{vecchia1985periodic, bloomfield1994periodic, lund1995climatological, tesfaye2004identification} tend to be PARMA; SARMA structures are useful in economics \citep{franses1994multivariate, franses2004periodic, hurd2007periodically}.  PARMA reviews are \cite{gardner1975characterization, lund1999modeling}, and \cite{gardner2006cyclostationarity}; statistical inference for PARMA models is studied in \cite{lund2000recursive, basawa2001large, basawa2004first, lund2006parsimonious, shao2004least}, and \cite{shao2006mixture}.  SARMA inference is addressed in \cite{chatfield1973box}.

\section{Methodology}

Our methods extend the work in \cite{jia2021count} with Gaussian transformations (copulas) to the periodic setting.  Let $\{ X_t \}$ denote the time series to be constructed, which takes values in the count support set $\{0, 1, 2, \ldots \}$.  Our construction works with a latent Gaussian series $\{ Z_t \}$ with zero mean and a unit variance at all times. Then $X_t$ is obtained from $Z_t$ via
\begin{equation}
\label{groundzero}
X_{dT+\nu}= F_{\nu}^{-1}\left( \Phi(Z_{dT+\nu}) \right),
\end{equation}
where $\Phi(\cdot)$ is the cumulative distribution function (CDF) of the standard normal distribution and $F_\nu(\cdot)$ is the desired marginal distribution for $X_t$ during season $\nu$. Here, $F_\nu^{-1}$ is the quantile function
\begin{equation}
F_{\nu}^{-1}(u) = \inf\left\{x: F_{\nu}(x) \geq u \right \}.
\label{eq:quantile}
\end{equation}
As \cite{jia2021count} shows, this construction leaves $X_{dT+\nu}$ with the marginal distribution $F_\nu$ for every $d$ and $\nu$. This model is very flexible: any marginal distribution can be achieved for any desired season $\nu$, even continuous ones. The marginal distribution $F_\nu$ can have the same form or be different for distinct seasons $\nu$. Any marginal distribution whatsoever can be achieved; when count distributions are desired, the quantile definition in (\ref{eq:quantile}) is the version of the inverse CDF that produces the desired marginals.

\subsection{Properties of the Model}

Toward ARMA and PARMA model order identification, if $\{ Z_t \}$ is an $m$-dependent series, then $Z_{t_1}$ and $Z_{t_2}$ are independent when $|t_1-t_2|>m$ since $\{ Z_t \}$ is Gaussian.  By (\ref{groundzero}), $X_{t_1}$ and $X_{t_2}$ are also independent and $\{ X_t \}$ is also $m$-dependent.  From the characterization of stationary moving averages (Proposition 3.2.1 in \cite{Brockwell_Davis_1991}) and periodic moving-averages in \cite{Shao_Lund_2004}, we see that if $\{ Z_t \}$ is a periodic moving average of order $q$, then $\{ X_t \}$ is also a periodic moving average of order $q$.   Unfortunately, analogous results for autoregressions do not hold. For example, if $\{ Z_t \}$ is a periodic first order autoregression, $\{ X_t \}$ may not be a periodic autoregression of any order \citep{jia2021count}.  

We now derive the covariance structure of $\{ X_t \}$ via Hermite expansions.  Let $\gamma_{X}(t,r) = \mbox{Cov}(X_t, X_r)$ be the covariance of $\{  X_t \}$ at times $t$ and $r$, where $r \leq t$.  Then $\gamma_X(t,r)$ can be related to $\rho_Z(t,r)$ via Hermite expansions.  To do this, let $G_\nu(x)=F^{-1}_{\nu}(\Phi(x))$ and write the Hermite expansion of $G_\nu(\cdot)$ as
\begin{equation}
G_\nu(z) = g_0(\nu) +\sum_{k=1}^{\infty}g_k(\nu)H_k(z).
\end{equation}
Here, $g_k(\nu)$ is the $k$th Hermite coefficient for season $\nu$, whose calculation is described below, and $H_k(z)$ is the $k$th Hermite polynomial defined by
\begin{equation}
H_k(z) = (-1)^ke^{z^2/2}\dfrac{d^k}{dz^k}\left( e^{-z^2/2} \right). 
\label{eq:Hermite}
\end{equation}
The first three Hermite polynomials are $H_0(x) \equiv 1$, $H_1(x) = x$, and $H_2(x) = x^2-1$. Higher order polynomials can be found via the recursion $H_k(x)=xH_{k-1}(x) - H_{k-1}^\prime (x)$, which follows from (\ref{eq:Hermite}).

The polynomials $H_k$ and $H_j$ are orthogonal with respective to the standard Gaussian measure if $k \neq j$: $E[ H_k(Z)H_j(Z)]= 0$ for a standard normal $Z$ unless $j=k$ (in which case $E[H_k(Z)^2]= k!$).  The Hermite coefficients are computed from
\begin{equation}
\label{gks}
g_k(\nu) = \dfrac{1}{k!}\int_{-\infty}^{\infty} G_{\nu}(t)H_k(t)\phi(t)dt, \quad k=1, 2, \ldots,
\end{equation}
where $\phi(t) = \Phi^\prime(t)= e^{-t^2/2}/\sqrt{2 \pi}$ is the standard normal density.

Lemma 2.1 in \cite{han2016correlation} shows that 	
\begin{equation}
\label{link_correlation}
\gamma_X(t,r) = \sum_{k=1}^{\infty}k!g_k(s(t))g_k(s(r))
\rho_Z(t,r)^k,
\end{equation}
where $s(t) = t - T \lfloor (t+1)/T \rfloor$ denotes the season corresponding to time $t$. Let $\sigma_X^2(t) = \gamma_X(t,t) = \sum_{k=1}^{\infty}k!g_k^2(s(t))$ denote the variance of $X_t$. Then the ACF of $\{ X_t \}$ is
\begin{equation}
\label{eq:link}
\rho_X(t,r)=\frac{\gamma_X(t,r)}{\sigma_X(t)\sigma_X(r)}
=\sum_{k=1}^{\infty}
\frac{k!g_k(s(t))g_k(s(r))}{\sigma_X(t)\sigma_X(r)}
\rho_Z(t,r)^k=\sum_{k=1}^{\infty}\ell_k
\rho_Z(t,r)^k:=L(\rho_Z(t,r)),
\end{equation}
which is a power series in $\rho_Z(t,r)$ with $k$th coefficient
\begin{equation}
\label{eq:link_coefficient}
\ell_k := 
\frac{k!g_k(s(t))g_k(s(r))}{\sigma_X(t)\sigma_X(r)}.
\end{equation}
\cite{jia2021count} call $L(\cdot)$ a link function and $\ell_k$ a link coefficient. When $\{ Z_t \}$ is stationary and $F_\nu$ does not depend on $\nu$, they show that $L(0)=0$, $L(1)=1$, and $L(-1)=\mbox{Corr} (G(Z_0),G(-Z_0))$.  It is not true that $L(-1)=-1$ in any case nor is $L(1)=1$ in the periodic case; indeed, stationary or periodically stationary count processes with arbitrarily positive or negative correlations may not exist.  For example, the pair $(Z, -Z)$, where $Z$ is standard normal has correlation -1, but two Poisson random variables, both having mean $\lambda$, whose correlation is -1, do not exist.

The model produces the most flexible correlation structures possible in a pairwise sense.  Specifically, consider two distinct seasons $\nu_1$ and $\nu_2$ and suppose that $F_{\nu_1}$ and $F_{\nu_2}$ are the corresponding marginal distributions for these seasons.  Then Theorems 2.1 and 2.5 in \cite{whitt1976bivariate} show that the bivariate random pair $(X_{\nu_1}, X_{\nu_2})$ having the marginal distributions $F_{\nu_1}$ and $F_{\nu_2}$, respectively, with the largest correlation has form $X_{\nu_1}= F^{-1}_{\nu_1}(U)$ and $X_{\nu_2}= F_{\nu_2}^{-1}(U)$, where $U$ is a uniform[0,1] random variable.  To achieve the largest correlation, one simply takes $\{ Z_t \}$ to have unit correlation at these times; that is, take $Z_{\nu_1}= Z_{\nu_2}$. Since $\Phi(Z_{\nu_1})$ is uniform[0,1], the claim follows.  For negative correlations, the same results in \cite{whitt1976bivariate} also show that the most negatively correlated pair that can be produced has the form $X_{\nu_1}=F_{\nu_1}^{-1}(U)$ and $X_{\nu_2}=F_{\nu_2}^{-1}(1-U)$.   This is produced with a Gaussian series having $\mbox{Corr}(Z_{\nu_1}, Z_{\nu_2})=-1$, which is obtained by selecting $Z_{\nu_2}=-Z_{\nu_1}$.  Then $\Phi(Z_{\nu_1})$ is again uniform[0,1] and $\Phi(Z_{\nu_2})=\Phi(-Z_{\nu_1})=1-\Phi(Z_{\nu_1})$, since $\Phi(-x)=1-\Phi(x)$ for all real $x$. 

The previous paragraph implies that one cannot construct more general autocorrelation functions for count series than what has been constructed above --- they do not exist.  Negatively correlated count series do arise \citep{kachour2009first, livsey2018multivariate, jia2021count} and can be described with this model class.   In the stationary case where the marginal distribution $F_\nu$ is constant over all seasons $\nu$, a series $\{ X_t \}$ with $\mbox{Cov}(X_t,X_{t+h})=1$ for all $h$ is achieved by taking $Z_t \equiv Z$, where $Z$ is standard normal.  This unit correlation property will not carry over to our periodic setting.  For example, a random pair $(X_{\nu_1}, X_{\nu_2})$ having a Poisson marginal with mean $\lambda_{\nu_1}$ during season $\nu_1$ and a Poisson marginal with mean $\lambda_{\nu_2}$ during season $\nu_2$ with a unit correlation do not exist when $\lambda_{\nu_1} \ne \lambda_{\nu_2}$.  This said, the model can produce any correlation structures within ``the range of achievable correlations".  As such, the model class here is quite flexible. 

The amount of autocorrelation that $\{ X_t \}$ inherits from $\{ Z_t \}$ is now discussed.  An implication of the result below, which establishes monotonicity of the link function by showing that its derivative is positive, is that the larger the autocorrelations are in $\{ Z_t \}$, the larger the autocorrelations are in $\{ X_t \}$.  We state the result below and prove it in the Appendix.

\noindent {\bf Proposition 3.1:} {\it For a fixed $t$ and $r$, let $L(\cdot)$ denote the link function in (\ref{eq:link}). Then for $u \in (-1, 1)$, the derivative of the link is positive and has form}
\begin{equation}
\label{e:link-derivative-again}
L'(u) = \frac{
\sum_{j_1=0}^\infty \sum_{j_2=0}^\infty 
e^{-\frac{1}{2(1-u^2)}\left[ \Phi^{-1}(C_{j_1}(s(t))^2 + \Phi^{-1}(C_{j_2}(s(r))^2 - 2 u \Phi^{-1}(C_{j_1}(s(t)) \Phi^{-1}(C_{j_2}(s(r))\right]}}
{\sqrt{\sigma_X(t)\sigma_{X}(r)} 2\pi \sqrt{1-u^2}}.
\end{equation}
Here, 
\begin{equation}
\label{eq:c_n}
C_{j}(\nu)=\mathbb{P}[ X_\nu \leq j]
\end{equation} 
denotes the cumulative probabilities of $X_\nu$ at season $\nu$.

\subsection{Calculation and Properties of the Hermite Coefficients}

An important numerical task entails calculating $g_k(\nu)$, which only depends on $F_{\nu}(\cdot)$ by (\ref{gks}).  To do this, rewrite $G_\nu(z)$ in the form
\begin{equation}
G_\nu(z)=\sum_{j=0}^{\infty}
j \mathbb{1}_
{ [C_{j-1}(\nu) \leq  \Phi^{-1}(z) < C_{j}(\nu) ]}
=\sum_{j=1}^{\infty}j
\mathbb{1}_{\left[ \Phi^{-1}(C_{j-1}(\nu)),\Phi^{-1}(C_j(\nu)) \right)}(z),
\end{equation}
where the convention $C_{-1}=0$ is made. We also take $\Phi^{-1}(0)= -\infty$ and $\Phi^{-1}(1)=\infty$. Then for $k \geq 1$, integration by parts yields
\begin{eqnarray*}
g_k(\nu) &=& \frac{1}{k!}\sum_{j=0}^{\infty}n\mathbb{E}\left[ \mathbb{1}_{\left[ \Phi^{-1}(C_{j-1}(\nu)),\Phi^{-1}(C_j(\nu)) \right)}(Z_0)H_k(Z_0) \right]\\
&=&\frac{1}{k!}\sum_{j=0}^{\infty}\frac{j}{\sqrt{2\pi}}\int_{\Phi^{-1}(C_{j-1}(\nu))}^{\Phi^{-1}(C_{j}(\nu))}H_k(z)e^{-z^2/2}dz\\
&=&\frac{1}{k!}\sum_{j=0}^{\infty}\frac{j}{\sqrt{2\pi}}\int_{\Phi^{-1}(C_{j-1}(\nu))}^{\Phi^{-1}(C_{j}(\nu))}(-1)^k\left( \frac{d^k}{dz^k}e^{-z^2/2} \right)dz\\
&=&\frac{1}{k!}\sum_{j=0}^{\infty}\frac{j}{\sqrt{2\pi}}(-1)^k\left( \frac{d^{k-1}}{dz^{k-1}}e^{-z^2/2} \right)\Bigg|_{z=\Phi^{-1}(C_{j-1}(\nu))}^{z=\Phi^{-1}(C_j(\nu))}\\
&=&\frac{1}{k!}\sum_{j=0}^{\infty}\frac{j}{\sqrt{2\pi}}(-1)e^{-z^2/2}H_{k-1}(z)\Bigg|_{z=\Phi^{-1}(C_{j-1}(\nu))}^{z=\Phi^{-1}(C_j(\nu))}.\\
\end{eqnarray*}
Simplifying this telescoping sum gives
\begin{equation}
\label{eq:g_k}
g_k(\nu) = \dfrac{1}{k!\sqrt{2\pi}}\sum_{j=0}^{\infty}
e^{-[\Phi^{-1}(C_{j}(\nu))]^2/2}H_{k-1}(\Phi^{-1}(C_{j}(\nu))).
\end{equation}
Notice that the summands in (\ref{eq:g_k}) are zero whenever $\Phi^{-1}(C_j(\nu))=\pm\infty$. Lemma 2.1 in \cite{jia2021count} shows that the expansion converges whenever $\mathbb{E}[ X_t^p ] < \infty$ for some $p>1$. This condition automatically holds for time series, which implicitly require a finite second moment. For count distributions with a finite support, $C_j(\nu)$ becomes unity for large $j$. For example, a Binomial marginal distribution with 7 trials is considered in our later application.  Here, the summation can be reduced to $j \in \{ 0, 1, \ldots, 7 \}$.  For count distributions on a countably infinite support, approximating (\ref{eq:g_k}) requires truncation of an infinite series.  This is usually not an issue:  numerically, $C_j(\nu)$ quickly converges to unity as $j \rightarrow \infty$ for light tailed distributions --- or equivalently, $e^{-\Phi^{-1}(C_{j}(\nu))^2/2}H_{k-1}(\Phi^{-1}(C_{j}(\nu)) \rightarrow 0$.  In addition to (\ref{eq:g_k}), $g_k(\nu)$ can also be approximated by Gaussian quadrature; see {\it gauss.quad.prob} in the package {\it statmod} in {\it R}. However, the approximation in (\ref{eq:g_k}) is more appealing in terms of simplicity and stability \citep{jia2021count}.

\section{Parameter Inference}

This section develops likelihood methods of inference for the model parameters via particle filtering and sequential Monte Carlo techniques.  With many count time series model classes, likelihoods are intractable \citep{davis2021count}.  Accordingly, researchers have defaulted to moment and composite likelihood techniques. However, if the count distributional structure truly matters, likelihood methods should ``feel" this structure and return superior parameter estimates. Gaussian pseudo-likelihood estimates, which are based only on the mean and autocovariance of the series, are developed in \cite{jia2021count} in the stationary case. \cite{jia2021count} presents an example where Gaussian pseudo-likelihood estimates perform well and an example where they perform poorly.

For notation, let $\boldsymbol{\theta}$ contain all parameters in the $T$ marginal distributions $F_1, \ldots, F_T$ and $\boldsymbol{\eta}$ contain all parameters governing the evolution of $\{ Z_t \}$.  The data $\{ x_1, x_2, \ldots, x_n \}$ denote our realization of the series.

The likelihood function $\mathcal{L}(\boldsymbol{\theta}, \boldsymbol{\eta})$ is simply a high dimensional multivariate normal probability. To see this, use (\ref{groundzero}) to get
\begin{equation}
\label{eq:likilihood}
\boldsymbol{\mathcal{L}}( \boldsymbol{\theta},\boldsymbol{\eta})
= \mathbb{P}(X_1=x_1, \cdots, X_n = x_n)
= \mathbb{P}\left( Z_1 \in (a_1,b_1], \cdots, Z_n \in (a_n,b_n] \right)
\end{equation}
for some numbers $\{ a_i \}_{i=1}^n$ and $\{ b_i \}_{i=1}^n$ (these are clarified below but are not important here). The covariance matrix of $(Z_1, \ldots , Z_n)$ only depends on $\boldsymbol{\eta}$, not on $\boldsymbol{\theta}$. Unfortunately, evaluating a high dimensional multivariate normal probability is numerically challenging for large $n$. While methods to handle this problem exist \citep{kazianka2010copula, kazianka2013approximate, bai2014efficient}, they often contain substantial estimation bias. 

An alternative approach, which is the one taken here, uses simulation methods to approximate the likelihood. General methods in this category include the quasi-Monte Carlo methods of \cite{genz2002comparison} and the prominent Geweke–Hajivassiliou–Keane (GHK) simulator of \cite{geweke1991efficient} and \cite{hajivassiliou1996simulation}. The performance of these methods, along with an additional ``data cloning" approach, are compared in \cite{han2020maximum}, where the author shows that estimators from these methods are similar, but that the GHK methods are much faster, having a numerical complexity as low as order $mn$. Here, $m$ is the pre-selected number of sample paths to be simulated (the number of particles). As we will subsequently see, the GHK simulator works quite well for large $m$.

\cite{jia2021count} propose a sequential importance sampling method that uses a modified GHK simulator. In essence, importance sampling is used to evaluate integrals by drawing samples from an alternative distribution and averaging their corresponding weights. Suppose that we seek to estimate $\int_{\mathcal{D}}f(\boldsymbol{x})d\boldsymbol{x}$.   Then
\[
\int_{\mathcal{D}}f(\boldsymbol{x})d\boldsymbol{x} =  \int_{\mathcal{D}}\dfrac{f(\boldsymbol{x})}
{q(\boldsymbol{x})}q(\boldsymbol{x})d\boldsymbol{x},
\]
where $f(\boldsymbol{x})/q(\boldsymbol{x})$ is called the weight and the proposed distribution $q$ is called the importance distribution. Without loss of generality, we assume that $q(\boldsymbol{x}) > 0$ whenever $\boldsymbol{x} \in \mathcal{D}$ and that $q(\boldsymbol{x})=0$ for $\boldsymbol{x} \in \mathcal{D}^c$. Then the importance sampling estimate of the integral is the law of large numbers justified average
\[
\int_{\mathcal{D}}\dfrac{f(\boldsymbol{x})}{q(\boldsymbol{x})}q(\boldsymbol{x})d\boldsymbol{x} \approx \dfrac{1}{m}\sum_{k=1}^{m}\dfrac{f(\boldsymbol{x}^{(k)})}{q(\boldsymbol{x}^{(k)})},
\]
where $\{\boldsymbol{x}^{(1)}, \ldots, \boldsymbol{x}^{(m)}\}$ are $m$ IID samples drawn from the proposed distribution $q$. With the notation $z_{1:n}=(z_1, \ldots, z_n)$, notice that the likelihood in  (\ref{eq:likilihood}) has form
\begin{equation}
\label{eq:likelihood_int}
\int_{ \{ z_t \in (a_t, b_t], t=1, \ldots, n \} } \boldsymbol{\phi}_{\boldsymbol{\eta}}\left( z_{1:n}\right)dz_1 \ldots dz_n = \int_{\{z_t\in (a_t, b_t], t=1, \ldots, n \}}\dfrac{\boldsymbol{\phi}_{\boldsymbol{\eta}}\left(z_{1:n}\right)}{q(z_{1:n})}q(z_{1:n})dz_1 \ldots dz_n. 
\end{equation}
Observe that $\{ a_i \}_{i=1}^n$ and $\{ b_i\}_{i=1}^n$ only depend on $\boldsymbol{\theta}$ and the data $\{x_1, x_2, \ldots, x_n\}$. Specifically,
\[
a_t = \Phi^{-1}(C_{x_t-1}(s(t))) \quad \mbox{and} \quad b_t = \Phi^{-1}(C_{x_t}(s(t))),
\]
where $C_{n}(v)$ is defined in (\ref{eq:c_n}) and $s(t)$ is the season at time $t$. Here, it is best to choose a proposed distribution $q$ such that 1) $q(z_{1:n}) > 0$ for $z_t \in (a_t, b_t]$ and $q(z_{1:n})=0$ otherwise; 2) the weight $\boldsymbol{\phi}_{\boldsymbol{\eta}} \left( z_{1:n}\right)/q(z_{1:n})$ is easy to compute; and 3) $\{ Z_t \}$ can be efficiently drawn from $q$. Our GHK simulator satisfies all three conditions.

To develop our GHK sampler further, we take advantage of the latent Gaussian structure in the PARMA or SARMA series $\{ Z_t \}$. In simple cases, $\{ Z_t \}$ may even be a Markov chain.  The GHK algorithm samples $Z_t$, depending on the its previous history $Z_{t-1}, \ldots, Z_1$ and $X_t$, from a truncated normal density. Specifically, let $p_{\boldsymbol{\eta}(t)} \left( z_t | z_{t-1}, \cdots, z_{1}, x_{t} \right)$ denote the truncated normal density of $Z_t$ given the history $Z_{t-1}, \ldots, Z_1$ and $X_t=x_t$. Then
\begin{equation}
\label{eq:p_density}
p_{\boldsymbol{\eta}(t)} \left( z_t | z_{t-1}, \ldots, z_{1}, x_{t} \right) = \dfrac{1}{r_t}
\left[ 
\dfrac{\phi(\frac{z_t-\hat{z}_t}{r_t})}{\Phi(\frac{b_t-\hat{z}_t}{r_t})-\Phi(\frac{a_t-\hat{z}_t}{r_t})}
\right], \quad a_t < z_t < b_t, 
\end{equation}
where $\hat{z}_t$ and $r_t$ are the one-step-ahead mean and standard deviation of $Z_t$ conditioned on $z_1, z_2, \ldots, z_n$. Again, $a_t$ and $b_t$ only depend on $x_t$. Here, we choose the importance sampling distribution 
\begin{equation}
q_{\boldsymbol{\eta}}(z_{1:n}|x_{1:n}) = p_{\boldsymbol{\eta}(1)}(z_1|x_1)
\prod_{t=2}^{n}p_{\boldsymbol{\eta}(t)}
\left( z_t | z_{t-1}, \ldots, z_{1}, x_{t} \right).
\end{equation}

Elaborating further, let $\mathcal{N}(\mu, \sigma^2; a, b)$ denote a normal random variable with mean $\mu$ and variance $\sigma^2$ that is known to lie in $(a,b]$, where $a < b$.  Then $X_1$ is first drawn from $\mathcal{N}(0, 1; a_1, b_1)$.  Thereafter, $X_2, X_3, \ldots, X_n$ are sequentially sampled from the distribution in (\ref{eq:p_density}). The proposed importance sampling distribution is efficient to sample, has the desired distributional support, and induces an explicit expression for the weights:
\[
\frac{\boldsymbol{\phi}_{\boldsymbol{\eta}(t)}\left( z_{1:n}\right)}{q_{\boldsymbol{\eta}(t)}(z_{1:n}|x_{1:n})} 
= 
\dfrac{p_{\boldsymbol{\eta}(1)}(z_1)}{p_{\boldsymbol{\eta}(1)}(z_1|x_1)}  \prod_{t=2}^{n}  \dfrac{ p_{\boldsymbol{\eta}(t)}\left(z_t\big|z_{t-1}, \ldots,z_{1} \right) }{p_{\boldsymbol{\eta}(t)}\left( z_t\big| z_{t-1}, \ldots, z_1, x_t \right)}. 
\]
Using (\ref{eq:p_density}) gives
\[
\dfrac{ p_{\boldsymbol{\eta}}\left(z_t\big|z_{t-1},\ldots,z_{1} \right) }{p_{\boldsymbol{\eta}}\left( z_t\big| z_{t-1},\ldots,z_1,x_t \right)}  = \Phi\left(\frac{b_t-\hat{z}_t}{r_t}\right)-\Phi\left(\frac{a_t-\hat{z}_t}{r_t}\right).
\]
Therefore,
\[
\frac{\boldsymbol{\phi}_{\boldsymbol{\eta}}\left( z_{1:n}\right)}{q(z_{1:n})} = 
\left[
\Phi\left( b_1 \right) -  \Phi\left( a_1 \right) \right] \prod_{t=2}^{n}\left[  \Phi\left( \dfrac{b_t-\hat{z}_t }{r_t}  \right)    - \Phi\left(\dfrac{a_t-\hat{z}_t }{r_t} \right)  \right].
\]
Define the initial weight $w_1 = \Phi(b_1) - \Phi(a_1)$.  We then recursively update the weights via 
\[
w_t = w_{t-1}\left[\Phi\left( \frac{b_t-\hat{z}_t }{r_t}  \right)    - \Phi\left(\frac{a_t-\hat{z}_t }{r_t} \right)\right]
\]
at time $t$ during the sequential sampling procedure. At the end of the sampling, we obtain 
\[
w_n = \frac{\boldsymbol{\phi}_{\boldsymbol{\eta}}( z_{1:n})}{q_{\boldsymbol{\eta}}(z_{1:n}|x_{1:n})}.
\]
In the classic GHK simulator, $\hat{Z}_t$ and $r_t^2$ are obtained from a Cholesky decomposition of the covariance matrix of $\{ Z_t \}$. Here, they are based on the PARMA or SARMA model for $\{ Z_t \}$.

The full sequential importance sampling procedure is summarized below.

\begin{itemize}

\item [1] Initialize the process by sampling $Z_1$ from the $\mathcal{N}(0,1; C_{x_1}(s(1)), C_{x_1}(s(1)))$ distribution. Define the weight $w_1$ by
\begin{equation}
w_1 = 
\Phi^{-1}(C_{x_{1}}(s(1))) - \Phi^{-1}(C_{x_{1}-1}(s(1)))
\end{equation}

\item [2] Now iterate steps 2 and 3 over $t=2, 3, \ldots, n$. Conditioned on $Z_1, \ldots, Z_{t-1}$, generate
\begin{equation}
Z_{t} \stackrel {{\cal D}} {=} 
\mathcal{N}\left( \hat{Z}_{t}, r_t;   \Phi^{-1}(C_{x_{t}}(s(t))), \Phi^{-1}(C_{x_{t}-1}(s(t))) \right).
\end{equation}
For example, in the PAR(1) model, $\hat{Z}_t = \phi(t) Z_{t-1}$ for $t \geq 1$, with the startup condition $\hat{Z}_1=0$; $r_t=1-\phi^2(t)$ for $t > 1$ with the startup $r_1=1$.

\item [3] Define the weight $w_t$ via
\begin{equation}
w_t = w_{t-1}~\left[ \Phi\left( \dfrac{\Phi^{-1}(C_{x_{t}}(s(t)))-\hat{Z}_t }{r_t}  \right)    - \Phi\left(\dfrac{\Phi^{-1}(C_{x_{t}-1}(s(t)))-\hat{Z}_t }{r_t} \right)  \right]
	\end{equation}
\end{itemize}

The above generates a fair draw of a single ``particle path" $\{ Z_t \}$ with the property that the $\{ X_t \}$ series generated from $\{ Z_t \}$ yields the observations $x_1, \ldots, x_n$.  Repeating this process $m$ independent times gives $m$ simulated process trajectories. Let $\{ {\bf Z}^{(1)}, \ldots, {\bf Z}^{(m)} \}$ be these trajectories and denote their corresponding weights at time $n$ by $\{ w_n^{(k)} \}_{k=1}^m$.

The importance sampling estimate is given by
\[
\hat{\mathcal{L}}^{\mbox{GHK}}\left( \boldsymbol{\theta},\boldsymbol{\eta} \right)=\dfrac{1}{m}\sum_{k=1}^{m}w_{n}^{(k)}.
\] 
A large $m$ provides more accurate estimation.

The popular ``L-BGSF-B" gradient step and search method is used to optimize the estimated likelihood $\hat{\mathcal{L}}^{\mbox{GHK}}(\boldsymbol{\theta}, \boldsymbol{\eta})$; other optimizers may also work.  However, $\hat{\mathcal{L}}^{\mbox{GHK}}(\boldsymbol{\theta},\boldsymbol{\eta})$ is ``noisy" due to the sampling. One popular fix to this smooths the estimated likelihood by generating a set of random quantities in the particle filtering through transformation and keeps them constant across the computations for different sets of parameters. This method, called common random numbers (CRNs), makes the simulated likelihood $\hat{\mathcal{L}}^{\mbox{GHK}}(\boldsymbol{\theta},\boldsymbol{\eta})$ relatively smooth in its parameters; see \cite{kleinman1999simulation} and \cite{glasserman1992some} for more on CRNs. In practice, the CRN point estimator behaves similarly to those for regular likelihoods; moreover, the Hessian-based covariance matrix, which is based on the derivative of $\hat{\mathcal{L}}^{\mbox{GHK}}(\boldsymbol{\theta}, \boldsymbol{\eta})$, behaves much better numerically when CRNs are used.  As the next section demonstrates, this procedure will yield standard errors that are very realistic.

Turning to model diagnostics, the probability integral transform (PIT) is used as a tool to evaluate model fitness. PIT methods, proposed in \cite{dawid1984present}, check the statistical consistency between probabilistic forecasts and the observations. Under the ideal scenario that the observations are drawn from the prediction distribution and the predictive distribution is continuous, PIT residuals are uniformly distributed over $[0, 1]$. PIT histograms tend to be $U$-shaped when the observations are over-dispersed. Unfortunately, the above themes do not hold for discrete count data. To remedy this, \cite{czado2009predictive} propose a nonrandomized PIT residual where uniformity still holds. Quantifying this, write the conditional cumulative distribution function of $X_t$ as
\begin{equation}
	 P_t(y):=\mathbb{P} \left( X_t \leq y | X_1 = x_1,\ldots,X_{t-1}=x_{t-1} \right), y \in \left\{ 0, 1, \ldots \right\} .
\end{equation}
Then the nonrandomized mean PIT residual is defined as $\bar{F}(u)= n^{-1}\sum_{t=1}^{n}F_t(u|x_t)$, where
\begin{equation}
	F_t(u|y)=\left\{\begin{array}{cl}
	0,&\hbox{if }u\leq P_t(y-1)\\
	\dfrac{u-P_t(y-1)}{P_t(y)-P_t(y-1)},&\hbox{if }P_t(y-1)<u<P_t(y)\\
	1,&\hbox{if }u\geq P_t(y)
	\end{array}
	\right..
\end{equation}
The quantity $P_t(y)$ can be approximated during the particle filtering algorithms; specifically,
\begin{equation}
	\hat{P}_t(y) = \sum_{i=0}^{y}w_{i,t}(\hat{Z}_t),
\end{equation}
where 
\[
w_{i,t}(z)=\Phi\left(\dfrac{\Phi^{-1}(C_i\left(s(t)\right))-z}{r_t}  \right)-\Phi\left(\dfrac{\Phi^{-1}(C_{i-1}\left(s(t)
\right))-z}{r_t} \right).
\]
The weight $w_{i,t}(z)$ can be obtained at time $t$ during the particle filtering algorithm. 

\section{Simulations}
This section presents a simulation study to evaluate the performance of our estimation methods. Periodic time series models often have a large number of parameters.  One way of consolidating these parameters into a parsimonious tally involves placing Fourier parametric constraints on the model parameters \citep{lund2006parsimonious, anderson2007fourier}, as is done below.   

\subsection{Poisson Marginals}
Our first simulation examines the classical Poisson count distribution with the PAR(1) $\{ Z_t \}$ in Example 2.1.  Here, $F_\nu$ is taken as a Poisson marginal with mean $\lambda(\nu)$, where the first-order Fourier constraint
\[
\lambda(\nu) = a_1 + a_2\cos\left( \dfrac{2\pi(\nu-a_3)}{T} \right)
\]
is imposed to consolidate the $T$ mean parameters into three.  Here, $|a_2| < a_1$ is imposed to keep $\lambda(\nu)$ non-negative.  The periods $T=10$ and $T=50$ are studied, the latter taken to roughly correspond to our future application to weekly rainy day counts. Our $\{ Z_t \}$ process obeys 
\[
Z_t = \phi(t)Z_{t-1} + \epsilon_t \sqrt{1-\phi(t)^2},
\]
with the AR coefficient $\phi(\nu)$ also being constrained by a first-order Fourier series that induces a causal model:
\begin{equation}
\label{seasonalphi}
\phi(\nu) = b_1 + b_2\cos\left( \dfrac{2\pi(\nu-b_3)}{T} \right).
\end{equation}
These specifications ensure that $\{ Z_t \}$ is a standard normal process ($E[Z_t] \equiv 0$ and $\mbox{Var}(Z_t) \equiv 1$). The parameters chosen must be legitimate in that $\lambda(\nu)$ must be positive for each $\nu$ and the PAR(1) model must be causal.  A six-parameter scheme that obeys these constraints is $a_1 = 10, a_2 = 5, a_3 = 5; b_1 = 0.5, b_2 = 0.2$, and $b_3 = 5$, which is now studied.

For each MLE optimization, $m=500$ independent particles are used along with series lengths of $n=100$ and $n=300$.  CRN techniques are used to the ensure that the likelihood is relatively smooth with respect to its parameters. This is an essential step --- see \cite{masarotto2017gaussian, han2018gckrig} for more on CRNs. Identifiability issues with the phase shift Fourier parametrizations arise since $a\cos(\pi/2 - b) = -a\cos(b - \pi/2)$; because of this, we impose $a_3, b_3 \in [0,T)$. Finally, The popular quasi-Newton method L-BFGS-B is implemented to optimize the likelihoods \citep{steenbergen2006maximum}.
	
Figures 1 shows boxplots of parameter estimators aggregated from $500$ independent series of various lengths and periods. The sample means of the parameter estimators are all close to their true values. When $T=50$ and $n=100$, there are only two complete cycles of data to estimate parameters from.  For standard errors of these parameters, Table \ref{PoissonPAR(1)} reports two values: 1) sample standard deviations of the parameter estimators over the 500 runs (denominator of 499), and 2) the average (over the 500 runs) of standard errors obtained by inverting the Hessian matrix at the maximum likelihood estimate for each run (denominator of 500).  Additional simulations (not shown here) with larger sample sizes with $T=50$ show that any biases recede as $n$ increases.  

\begin{figure}[hbt!]
\centering
\includegraphics[width=14cm]{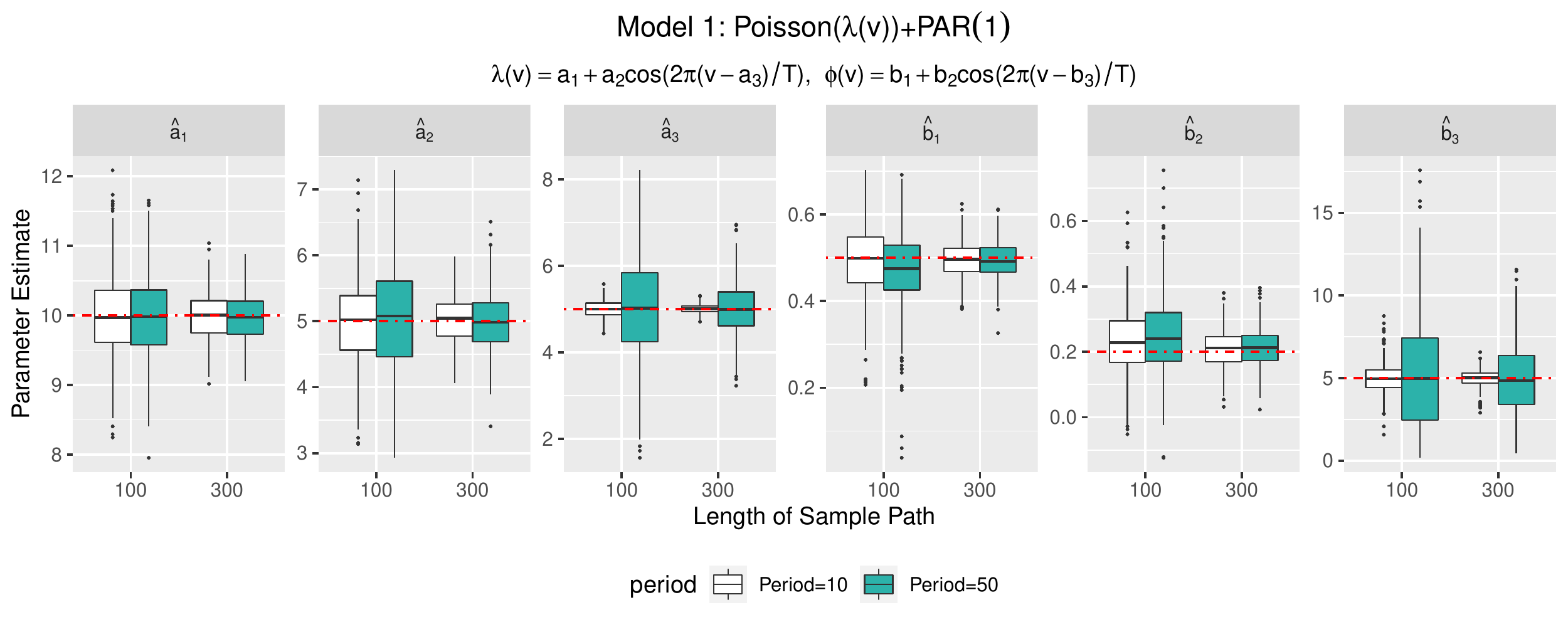}
\caption{Box plots of parameter estimators for a Poisson marginal distribution with a PAR(1) $\{ Z_t \}$. All estimators appear approximately unbiased --- the dashed lines demarcate true parameter values.}
\end{figure}

\begin{table}[hbt!]
	\centering
	\begin{threeparttable}
		\begin{tabular}{|c|c|c|c|c|c|c|c|c|}
			\headrow
			\hline
			\multicolumn{9}{|c|}{Model 1} \\
			n&T&&$ \hat{a_1} $&$ \hat{a_2} $&$ \hat{a_3} $&$ \hat{b_1} $&$ \hat{b_2} $&$ \hat{b_3} $\\ 
			&&mean&9.98375 &5.00087 &5.00443 &0.49213 &0.23474 &4.98639\\
			100&10&SD&0.60068 &0.63494 &0.18490 &0.07818 &0.09896 &0.90434\\ 
			&&$ \hat{E}(I'(\theta)^2) $&0.57199 &0.60025 &0.17375 &0.07720 &0.10282 &0.90060\\ \hline
			&&mean&9.98127 &5.05051 &5.01370 &0.46899 &0.25285 &5.24743\\
			100&50&SD&0.58975 &0.79520 &1.14622 &0.08861 &0.11470 &3.50616\\
			&&$ \hat{E}(I'(\theta)^2) $&0.59190 &0.76871 &1.11326 &0.08268 &0.11469 &4.51175\\ \hline
			&&mean&9.98652 &5.01761 &5.00714 &0.49608 &0.20812 &4.98741\\
			300&10&SD&0.33694 &0.36520 &0.10039 &0.04173 &0.05550 &0.48356\\
			&&$ \hat{E}(I'(\theta)^2) $&0.32824 &0.35012 &0.09958 &0.04151 &0.05569 &0.45228\\ \hline
			&&mean&9.97370 &4.98609 &5.00089 &0.49386 &0.20998 &4.89220\\
			300&50&SD&0.33449 &0.45038 &0.62028 &0.04188 &0.05726 &2.09052\\
			&&$ \hat{E}(I'(\theta)^2) $&0.34588 &0.45153 &0.65046 &0.04156 &0.05584 &2.33228\\ \hline
\end{tabular}
\caption{Standard errors for the parameter estimators for a Poisson marginal distribution with a PAR(1) $\{ Z_t \}$.  The results show the sample standard deviation (SD) of the parameter estimators from 500 independent series, and the average of the 500 standard errors obtained by inverting the Hessian matrix ($\hat{E}(I^\prime(\theta)^2)$) at the maximum likelihood estimate over these same runs.}
\label{PoissonPAR(1)}
	\end{threeparttable}
\end{table}

We next consider the same Poisson marginal case, but now change $\{ Z_t \}$ to the SAR(1) series in Example 2.2. The $a_1$, $a_2$, and $a_3$ chosen for this simulation are the same as above. The SAR(1) parameters chosen are $\phi=0.5$ and $\alpha=0.3$.  Figure 2 shows boxplots of the parameter estimators akin to those in Figure 1.  The overall performance is again very good --- interpretations of the results are similar to those for the PAR(1) model above. Table \ref{PoissonSAR(1)} shows our two types of standard errors and again reveals nice agreement.

\begin{figure}[hbt!]
\centering
\includegraphics[width=14cm]{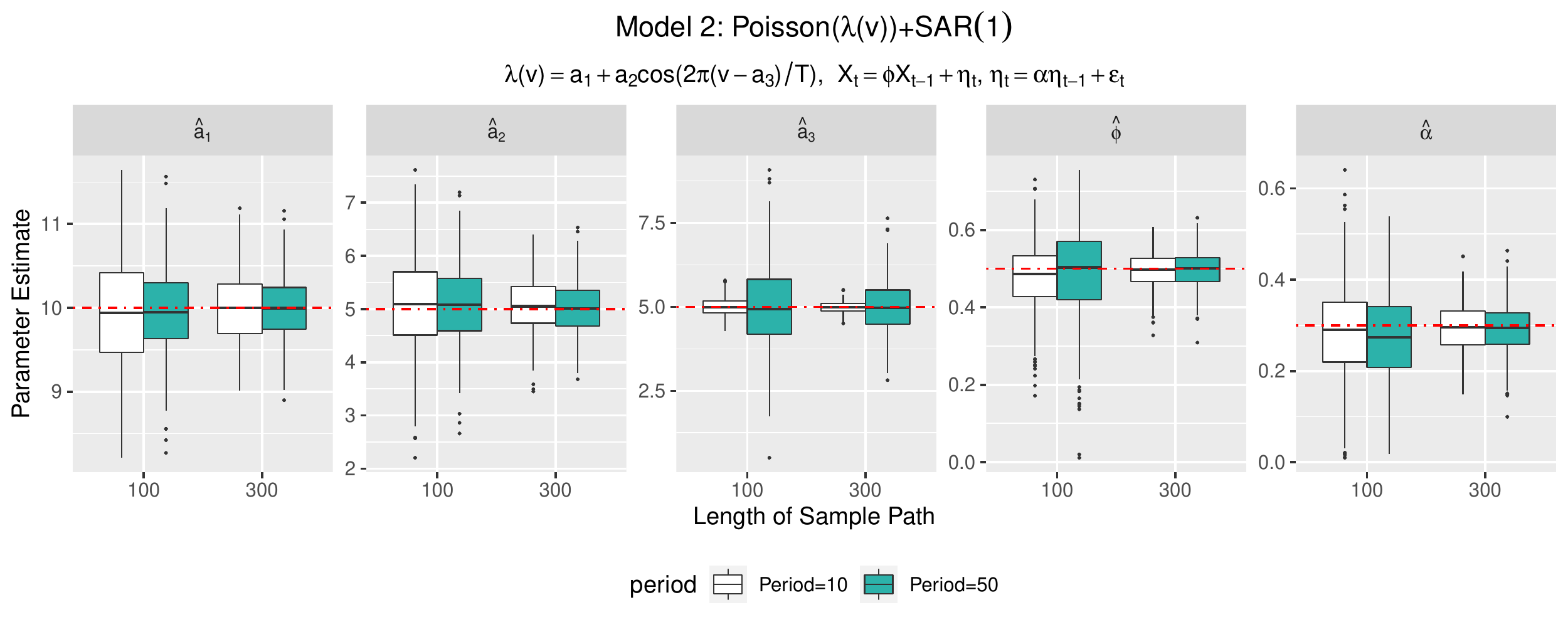}
\caption{Box plots of parameter estimators for the Poisson marginal distribution with a SAR(1) $\{ Z_t\}$. All estimators appear approximately unbiased --- the dashed lines demarcate true parameter values.}
\end{figure}

\begin{table}[hbt!]
\centering
	\begin{threeparttable}
\begin{tabular}{|c|c|c|c|c|c|c|c|}
			\headrow
			\hline
			\multicolumn{8}{|c|}{Model 2} \\
			n&T&&$ \hat{a_1} $&$ \hat{a_2} $&$ \hat{a_3} $&$ \phi $&$ \alpha $\\ 
			&&mean&9.94650 &5.07987 &4.99206 &0.47876 &0.28482\\ 
			100&10&SD&0.66096 &0.87087 &0.26175 &0.08455 &0.10242\\ 
			&&$ \hat{E}(I'(\theta)^2) $&0.65459 &0.80021 &0.25874 &0.08242 &0.09947\\ \hline
			&&mean&9.96430 &5.07165 &4.99623 &0.49003 &0.27121\\
			100&50&SD&0.51019 &0.71742 &1.16749 &0.11527 &0.09630\\
			&&$ \hat{E}(I'(\theta)^2) $&0.50809 &0.68786 &1.10774 &0.10345 &0.09874\\ \hline
			&&mean&9.99120 &5.06420 &4.99278 &0.49535 &0.29260\\
			300&10&SD&0.41347 &0.50127 &0.15764 &0.04399 &0.05155\\
			&&$ \hat{E}(I'(\theta)^2) $&0.40803 &0.50201 &0.15827 &0.04269 &0.05426\\ \hline
			&&mean&9.99133 &5.01493 &5.01603 &0.49874 &0.29189\\
			300&50&SD&0.37005 &0.49076 &0.78444 &0.04674 &0.05392\\
			&&$ \hat{E}(I'(\theta)^2) $&0.36755 &0.49687 &0.79666 &0.04543 &0.05403\\ \hline
		\end{tabular}
\caption{Standard errors for the parameter estimators for the Poisson marginal distribution with a SAR(1) $\{ Z_t \}$.  The results show the sample standard deviation (SD) of the parameter estimators from 500 independent series, and the average of the 500 standard errors obtained by inverting the Hessian matrix ($\hat{E}(I^\prime(\theta)^2)$) at the maximum likelihood estimate over these same runs.}
        \label{PoissonSAR(1)}
	\end{threeparttable}
\end{table}

\subsection{A Markov Chain Induced Marginal Distribution}	

Another marginal distribution that we consider is derived from a two-state Markov chain (TSMC) model.  This distribution will fit our weekly rainy day counts well in the next section.  Consider a Markov transition matrix ${\bf Q}$ on two states with form 
\[
{\bf Q}= 
\left[ \begin{array}{cc}
\alpha   & 1- \alpha \\
1-\beta & \beta     \\
\end{array}
\right] .
\]
Here, $\alpha \in (0,1)$ is interpreted as the probability that day $t+1$ is dry given that day $t$ is dry; analogously, $\beta \in (0,1)$ is the probability that day $t+1$ is rainy given that day $t$ is rainy.   Let $\{ M_t \}_{t=0}^7$ be a Markov chain with these transition probabilities.  The marginal distribution that we consider for $\{ X_t \}$ has the form
\[
\mathbb{P}( X_t =k ) = \mathbb{P}_{M_0} \left( \sum_{t=1}^7 M_t = k \right), \quad k \in \{ 0, 1, 2, 3, 4, 5, 6, 7 \},
\]
where $M_0 \in \{ 0, 1 \}$.  Here, $M_0=0$ signifies that the day before the week started was dry and $M_0=1$ signifies that the day before the week started was rainy. This marginal distribution, while difficult to derive in explicit form, allows for dependence in the day-to-day rain values, improving on a Binomial model with seven trials that models successive days as independent.  
	
It is not easy to derive an explicit form for the distribution of $F$; however, it can be built up numerically by allowing the number of days in a week to be a variable $L$ and recursing on it:
\begin{eqnarray}
\mathbb{P}_0 \left( \sum_{t=1}^L M_t = k \right) &=& 
(1-\alpha) \mathbb{P}_1 \left( \sum_{t=1}^{L-1} M_t = k-1 \right) + 
\alpha     \mathbb{P}_0 \left( \sum_{t=1}^{L-1} M_t = k \right); \\
\mathbb{P}_1 \left( \sum_{t=1}^L M_t = k \right) &=& 
\beta  \mathbb{P}_1 \left( \sum_{t=1}^{L-1} M_t = k-1 \right) + 
(1-\beta) \mathbb{P}_0 \left( \sum_{t=1}^{L-1} M_t  = k  \right).
\end{eqnarray}
These recursions start with probabilities for a one day week: $\mathbb{P}_0(M_t=1)=1-\alpha$; $\mathbb{P}_0(M_t=1)=\alpha$; $\mathbb{P}_1(M_t=1)=\beta$, and $\mathbb{P}_1(M_t=0)=1-\beta$. We take the initial state of the chain to be random with the stationary distribution
\[
\mathbb{P}(M_0=0)= \frac{1-\alpha}{2-\alpha-\beta}; \quad \mathbb{P}(M_0=1)=\frac{1-\beta}{2-\alpha-\beta}.
\]

To allow for periodicites in the above TSMC structure, we parametrize $\alpha$ and $\beta$ as short Fourier series again:
\[
\alpha(\nu)=a_1 + a_2\cos\left( \dfrac{2\pi(\nu-a_3)}{T} \right); \quad
\beta(\nu)=b_1 + b_2\cos\left( \dfrac{2\pi(\nu-b_3)}{T} \right).
\]

\begin{figure}[hbt!]
\centering
\includegraphics[width=14cm]{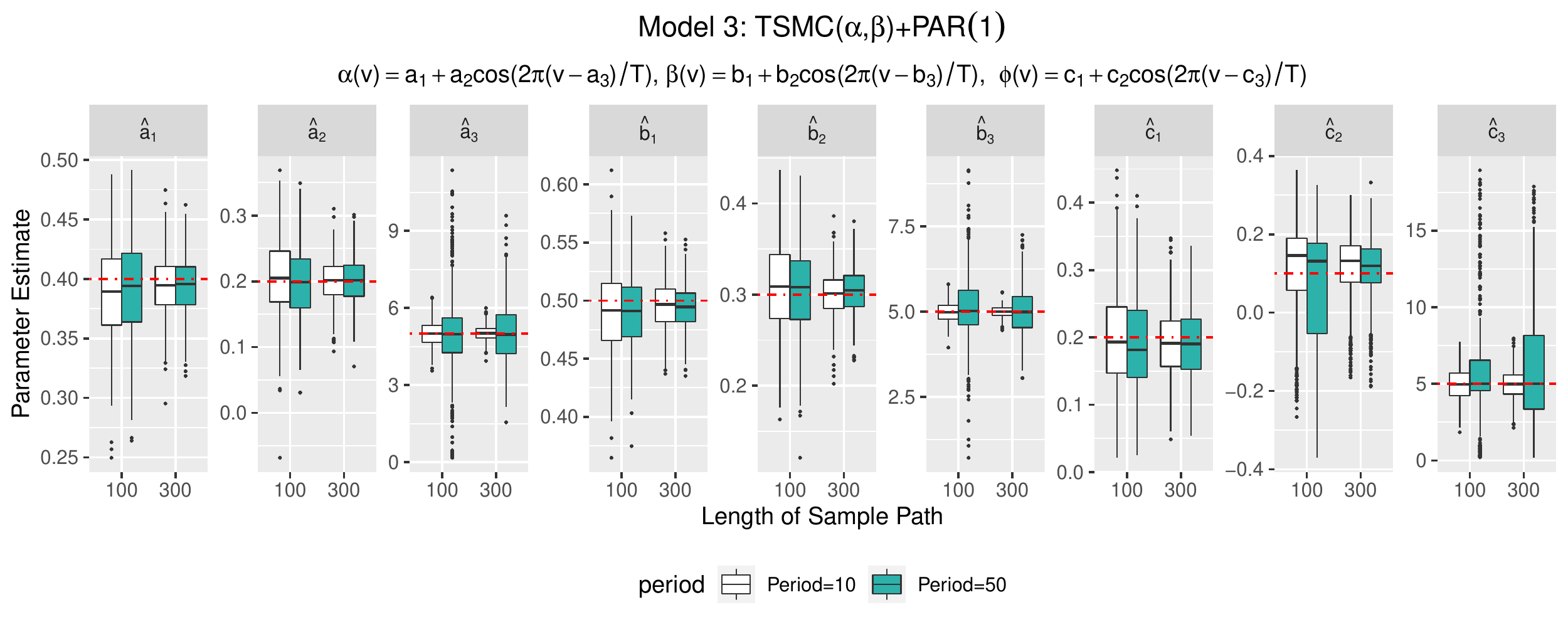}
\caption{Box plots of parameter estimators for the TSMC marginal distribution with a PAR(1) $\{ Z_t \}$. All estimators appear approximately unbiased --- the dashed lines demarcate true parameter values.}
\label{fig:TSMC+PAR(1)}
\end{figure}

Our first TSMC simulation considers the PAR(1) $\{ Z_t \}$ in (\ref{seasonalphi}).  This is a nine parameter model. The parameter values considered are $a_1 = 0.4, a_2 = 0.2, a_3 = 5; b_1 = 0.5, b_2 = 0.2, b_3 = 0.3, c_1 = 0.2, c_2 = 0.1$, and $c_3 = 5$, which induce a causal $\{ Z_t \}$ and legitimate Markov chain transitions (all transitions have non-negative probabilities).   Figure \ref{fig:TSMC+PAR(1)} shows boxplots of the estimated parameters over 500 independent series of various lengths and periods.  Table \ref{tab:TSMC+PAR(1)} shows standard errors computed from the two methods previously described. For the most part, the results are satisfying.  Some of the phase shift parameter's "Hessian inverted" standard errors are larger than the sample standard deviation standard errors.   The phase shift parameter is the argument where its associated cosine wave is maximal and lies in $[0,T]$. Because of the larger support set, this parameter will naturally have more variability than say parameters supported in (-1,1).  Also, when $n=100$ and $T=50$, there are only two complete cycles from which to estimate the location of this maximum --- this will be statistically difficult.  Additional simulations (not reported) show that these discrepancies recede as the sample size gets larger.

\begin{table}[hbt!]
	\centering
	\begin{threeparttable}
		\begin{tabular}{|c|c|c|c|c|c|c|c|c|c|c|c|}
			\headrow
			\hline
			\multicolumn{12}{|c|}{Model 3} \\
			n&T&&$ \hat{a_1} $&$ \hat{a_2} $&$ \hat{a_3} $&$ \hat{b_1} $&$ \hat{b_2} $&$ \hat{b_3} $&$ \hat{c_1} $&$ \hat{c_2} $&$ \hat{c_3} $\\
			&&mean&0.389 &0.205 &4.997 &0.490 &0.307 &4.985 &0.199 &0.105 &4.963\\
			100&10&SD&0.041 &0.060 &0.472 &0.036 &0.048 &0.295 &0.076 &0.134 &1.221\\ 
			&&$ \hat{E}(I'(\theta)^2) $&0.043 &0.059 &0.550 &0.035 &0.045 &0.280 &0.108 &0.178 &2.275\\ \hline
			&&mean&0.391 &0.199 &5.009 &0.490 &0.302 &5.109 &0.192 &0.072 &5.994\\
			100&50&SD&0.040 &0.057 &1.876 &0.032 &0.050 &1.137 &0.072 &0.153 &4.153\\
			&&$ \hat{E}(I'(\theta)^2) $&0.044 &0.065 &3.210& 0.036& 0.049& 1.603& 0.112& 0.195& 14.718\\ \hline
			&&mean&0.394& 0.200& 5.013& 0.496& 0.301& 5.002& 0.194& 0.114& 4.969\\
			300&10&SD&0.024& 0.033& 0.306& 0.021& 0.026& 0.159& 0.052& 0.084& 1.095\\
			&&$ \hat{E}(I'(\theta)^2) $&0.025& 0.033& 0.291& 0.020& 0.026& 0.159& 0.060& 0.087& 1.376\\ \hline
			&&mean&0.395& 0.201& 4.989& 0.495& 0.304& 5.017& 0.191& 0.108& 5.989\\
			300&50&SD&0.024& 0.035& 1.269& 0.020& 0.026& 0.733& 0.053 &0.085& 4.130\\
			&&$ \hat{E}(I'(\theta)^2) $&0.025& 0.034& 1.506& 0.020& 0.027& 0.823& 0.060& 0.097& 8.054\\ \hline
		\end{tabular}
\caption{Standard errors for the parameter estimators for the TSMC marginal distribution with a PAR(1) $\{ Z_t \}$.  The results show the sample standard deviation (SD) of the parameter estimators from 500 independent series, and the average of the 500 standard errors obtained by inverting the Hessian matrix ($\hat{E}(I^\prime(\theta)^2)$) at the maximum likelihood estimate over these same runs.}
		\label{tab:TSMC+PAR(1)}
	\end{threeparttable}
\end{table}

\begin{figure}[hbt!]
\includegraphics[width=14cm]{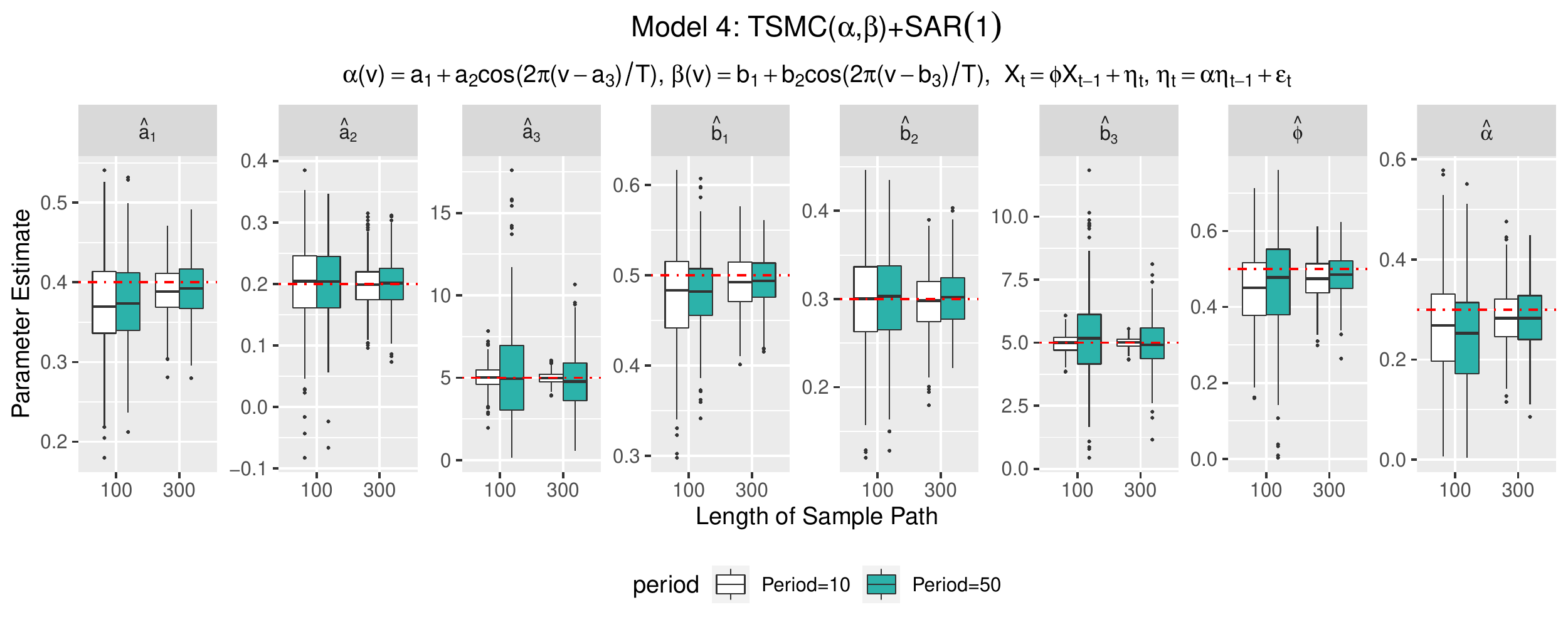}
\caption{Box plots of parameter estimators for the TSMC marginal distribution with a SAR(1) $\{ Z_t \}$. All estimators appear approximately unbiased --- the dashed lines demarcate true parameter values.}
\label{fig:TSMC+SAR(1)}
\end{figure}

Finally, we consider the TSMC marginal distribution with a SAR(1) $\{ Z_t \}$.  Figure \ref{fig:TSMC+SAR(1)} shows boxplots of the estimated parameters over 500 independent series of various lengths and periods.  Table \ref{tab:TSMC+SAR(1)} shows standard errors computed by our two methods. Again, the performance is good --- the interpretation of the results is analogous to that given before.

\begin{table}[hbt!]
\centering
	\begin{threeparttable}
		\begin{tabular}{|c|c|c|c|c|c|c|c|c|c|c|c|}
			\headrow
			\hline
			\multicolumn{11}{|c|}{Model 4} \\
            n&T&&$ \hat{a_1} $&$ \hat{a_2} $&$ \hat{a_3} $&$ \hat{b_1} $&$ \hat{b_2} $&$ \hat{b_3} $&$ \phi $&$ \alpha $\\ \hline
			&&mean&0.372 &0.200 &4.995 &0.480 &0.299 &5.010 &0.447 &0.260\\
			100&10&SD&0.061 &0.068 &0.605 &0.050 &0.056 &0.356 &0.107 &0.111\\ 
			&&$ \hat{E}(I'(\theta)^2) $&0.055 &0.063 &0.624 &0.046 &0.053 &0.335 &0.102 &0.109\\ \hline
			&&mean&0.378 &0.199 &4.890 &0.482 &0.299 &5.106 &0.469 &0.250\\
			100&50&SD&0.049 &0.060 &2.807 &0.044 &0.048 &1.564 &0.123 &0.104\\
			&&$ \hat{E}(I'(\theta)^2) $&0.049 &0.062 &3.602 &0.040 &0.050 &1.572 &0.121 &0.107\\ \hline
			&&mean&0.388 &0.201 &5.006 &0.493 &0.302 &5.004 &0.477 &0.289\\
			300&10&SD&0.033 &0.038 &0.317 &0.030 &0.032 &0.202 &0.056 &0.063\\
			&&$ \hat{E}(I'(\theta)^2) $&0.033 &0.038 &0.335 &0.028 &0.032 &0.193 &0.054 &0.060\\ \hline
			&&mean&0.390 &0.199 &5.030 &0.492 &0.302 &5.016 &0.483 &0.281\\
			300&50&SD&0.033 &0.040 &1.689 &0.027 &0.031 &1.061 &0.055 &0.059\\
			&&$ \hat{E}(I'(\theta)^2) $&0.032 &0.038 &1.706 &0.027 &0.032 &0.960 &0.056 &0.060\\ \hline
		\end{tabular}
\caption{Standard errors for the parameter estimators for the TSMC marginal distribution with a SAR(1) $\{ Z_t \}$.  The results show the sample standard deviation (SD) of the parameter estimators from 500 independent series, and the average of the 500 standard errors obtained by inverting the Hessian matrix ($\hat{E}(I^\prime(\theta)^2)$) at the maximum likelihood estimate over these same runs.}
		\label{tab:TSMC+SAR(1)}
	\end{threeparttable}
\end{table}

\section{Application}

\begin{figure}[h!]
\centering
\includegraphics[width=14cm]{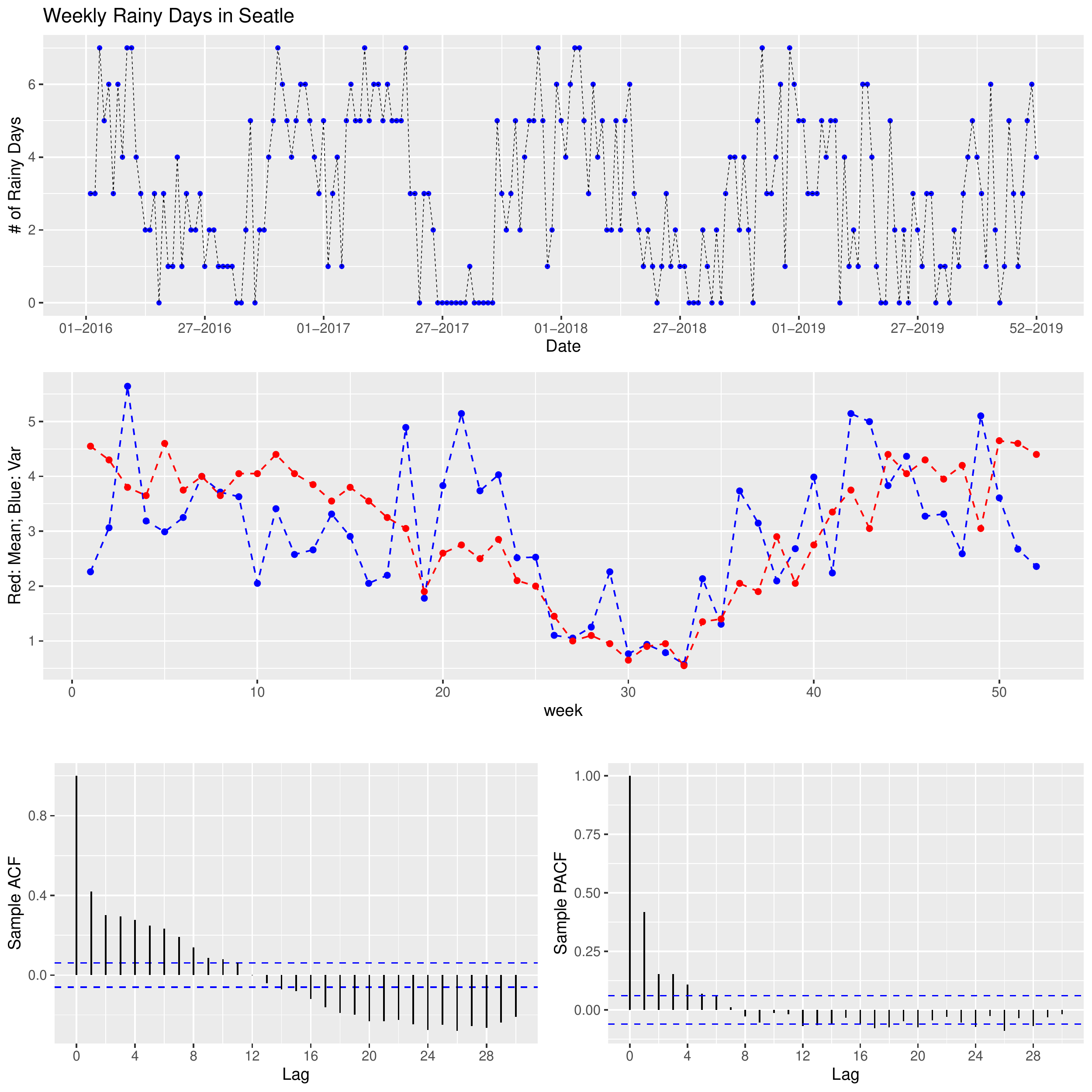}
\caption{Top: The Seattle weekly rainy day counts from 2016-2019 only; Middle: Weekly sample means and variance for the rainy day counts from 2000-2019; Bottom left and bottom right: Sample ACF and PACF of all observations.}
\label{fig:dataspecs}
\end{figure}

This section applies our techniques to a series of weekly rainy day counts in the Seattle, Washington area recorded from 01-Jan-2000 to 31-Dec-2019. The data were collected at the Seattle Tacoma airport weather station and are available at \url{http://www.ncdc.noaa.gov}.  Here, any day receiving a non-zero amount of precipitation is counted as a rainy day. As such, $X_t \in \{0, 1, 2, 3, 4, 5, 6 , 7 \}$ for each $t$.  For convenience, we only analyze the first 364 days in a year, inducing a period of $T=52$ in the series.  Any inaccuracies incurred by neglecting these days is minimal.  Figure 3 summarizes our data.  The top plot graphs the weekly rainy day counts from the last four years of the series only (for visual clarity), from the first week in 2016 to the 52nd week in 2019.  One sees a clear seasonal cycle with summer weeks experiencing significantly less rain than winter weeks.  The middle plot in the figure displays the sample mean and variance of the weekly counts over the entire 20 year data period, aggregated by week of year. For example, the mean and variance for the 1st week in January are the sample means and variance (denominator of 19) over all 20 1st weeks occurring from 2000 to 2019. The sample mean and variance have roughly sinusoidal structures and are minimal during the summer months. The bottom plots in the figure show sample autocorrelations (ACF) and partial autocorrelations (PACF) of the series. The pattern in the ACF is indicative of a periodic mean in the series that has not been removed.

Several marginal distributions for this series merit exploration.  The binomial distribution with seven trials is a classic structure for such data.  However, this distribution is underdispersed (variance is smaller than the mean), which does not jibe with the data patterns seen in the middle plot of Figure \ref{fig:dataspecs}. Another distribution considered is the TSMC distribution of the last section. This distribution can be overdispersed, and as we will subsequently see, fits our series quite well.  A final marginal distribution considered is the generalized Poisson marginal truncated to the support set $\{ 0, 1, 2, 3, 4, 5, 6, 7 \}$. For clarity, the generalized Poisson marginal we use has distribution 
\begin{eqnarray*}
\mathbb{P}(Y=k)&=&\frac{e^{-(\lambda+\eta k)} \lambda(\lambda+\eta k)^{k-1}}{k!}, \quad k=0,1,\ldots; \\
\mathbb{E}(Y)&=&\mu=\frac{\lambda}{1-\eta}; \\
\mbox{Var}(Y)&=&\sigma^2=\frac{\lambda}{(1-\eta)^3}
\end{eqnarray*}
for a count variable $Y$, with $\lambda>0$ and $\eta \in [0,1)$.  When $\eta=0$, $Y$ is Poisson($\lambda)$ and is equi-dispersed.  First order Fourier cosine constraints are placed on the mean and variance pair $(\mu(\nu), \sigma^2(\nu))$ and then mapped back to parameter pair $(\lambda(\nu), \alpha(\nu))$.

For structures of $\{ Z_t \}$, we consider PAR(1), AR(1), and SAR(1) models (see Section 2).  The PAR(1) structure uses the first order Fourier cosine consolidation in (\ref{seasonalphi}) for $\{ Z_t \}$. The AR(1) $\{ Z_t \}$ is simply a standard AR(1) series with a unit variance.  The SAR(1) form for $\{ Z_t \}$ is the two parameter model in Example 2.2.  For parameters in the marginal distributions, the success probabilities are
\[
p(\nu)=a_1 + a_2\cos\left( \frac{2\pi(a_3-\nu)}{T} \right)
\]
in the binomial fits;
\[
\alpha(\nu)=a_1 + a_2\cos\left( \frac{2\pi(a_3-\nu)}{T} \right), \quad 
\beta(\nu)=b_1 + b_2\cos\left( \frac{2\pi(b_3-\nu)}{T} \right),
\]
for the TSMC fits, and 
\begin{eqnarray*}
\mu(\nu)     = a_1 + a_2\cos\left( \frac{2\pi(a_3-\nu)}{T} \right),&& \quad \sigma^2(\nu)=b_1 + b_2\cos\left( \frac{2\pi(b_3-\nu)}{T} \right), \\
\lambda(\nu) = 1 - \sqrt{\frac{\mu(\nu)}{\sigma^2(\nu)}}, && \quad \alpha(\nu) = \mu(\nu)\sqrt{\frac{\mu(\nu)}{\sigma^2(\nu)}}
\end{eqnarray*}
in the truncated generalized Poisson fit. 

Table \ref{Tab:AICBIC} displays BIC and AIC scores for various fitted $\{ Z_t \}$ structures and marginal distributions. The best marginal distribution is the TSMC; the truncated generalized Poisson marginal distribution is a close second.  The generalized Poisson marginal distribution is known to be a very flexible count time series model \citep{ver2007quasi} that fits many observed series well.  Of note is that an AR(1) latent $\{ Z_t \}$ is preferred to either a PAR(1) or SAR(1) structure.   This does not mean that the end fitted model is non-periodic; indeed, the parameters in the marginal distribution $F_\nu$ depend highly on the week of year $\nu$.   However, the seasonality in the PAR(1) and SAR(1) $\{ Z_t \}$ do not make an appreciable difference --- a stationary AR(1) $\{ Z_t \}$ is sufficient.

\begin{table}[h!]
	\centering
	\begin{threeparttable}
		\begin{tabular}{cccccc}
			\headrow
			Marginal Distribution & Model & WN &AR(1) & PAR(1) & SAR(1)\\
			Binomial&AIC&4278.628&{\bf 4227.775}&4229.708&4229.458 \\
			&BIC&4293.469&{\bf 4247.563}&4259.39&4254.193\\
			Two State Markov Chain (TSMC)&AIC&3888.114&{\bf 3853.589}&3856.244&3855.624\\
			&BIC&3917.796&{\bf 3888.218}&3900.766&3895.200\\
			Truncated Overdispersed Poisson&AIC&3885.032&{\bf 3853.840}&3856.995&3855.672\\
			&BIC&3914.714&{\bf 3888.469}&3901.518&3895.248\\
			\hline  
		\end{tabular}
	\end{threeparttable}
\caption{AIC and BIC statistics for models with binomial, TSMC, and truncated Poisson marginal distributions. The lowest AIC/BIC for each marginal distribution are bolded. The TSMC marginal distribution with an AR(1) $\{ Z_t \}$ is judged optimal.}
	\label{Tab:AICBIC}
\end{table}

Table \ref{Paraest} shows the estimated parameters in the fitted model.  Based on asymptotic normality, which is expected but has not been proven, all parameters except $b_3$ appear to be significantly non-zero.   A zero $b_3$ is plausible: $b_3=0$ implies that the maximal variability of the weekly rainy day counts start at the beginning of the calendar year, which roughly corresponds to the meteorological height of winter.  Standard errors were estimated by inverting the Fisher information matrix at the likelihood estimates.  For completeness, Table \ref{paraest2} shows parameter estimates and standard errors for the truncated generalized Poisson fit.  The interpretation of these results are similar to those above.

\begin{table}[h!]
\caption{Estimates and standard errors of the TSMC AR(1) model. The L-BFGS-B algorithm was used to optimize particle filtering likelihoods.}
	\centering
	\begin{threeparttable}
		\begin{tabular}{cccccccc}
			\headrow
			Parameters&$ a_1 $&$ a_2 $&$ a_3 $&$ b_1 $&$ b_2 $&$ b_3 $&$ \phi $\\ \hline
			Point Estimates&0.737 &-0.163 & 4.687 & 0.648&  0.132 & 1.660 & 0.198\\
			Standard Error&0.011& 0.014& 0.674 &0.013 &0.018 &1.039& 0.032\\
			\hline  
		\end{tabular}
	\end{threeparttable}
	\label{Paraest}
\end{table}

\begin{table}[h!]
	\caption{Estimates and standard errors of the generalized Poisson-AR(1) fit. The L-BFGS-B algorithm is used to optimize particle filtering likelihoods.}
	\centering
	\begin{threeparttable}
		\begin{tabular}{cccccccc}
			\headrow
			Parameters&$ a_1 $&$ a_2 $&$ a_3 $&$ b_1 $&$ b_2 $&$ b_3 $&$ \phi $\\ \hline
			Point Estimates&3.999 &2.975 &3.977 &8.155 &6.926 &3.955& 0.195\\
			Standard Error&0.263 &0.298 &0.448& 1.586& 1.685 &0.926& 0.033\\
			\hline  
		\end{tabular}
	\end{threeparttable}
	\label{paraest2}
\end{table}

\begin{figure}[h!]
	\centering
	\includegraphics[width=14cm]{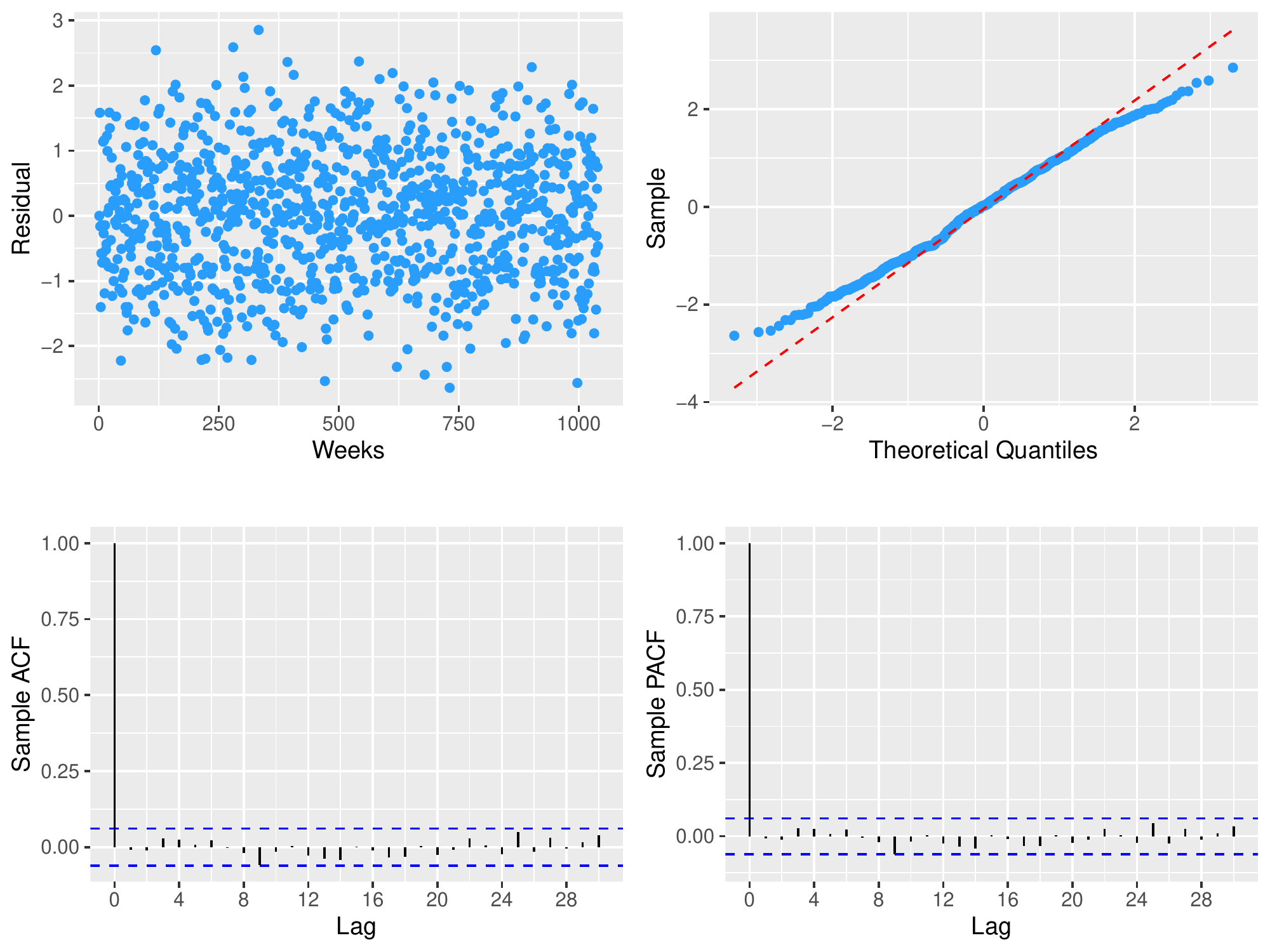}
\caption{Top left: TSMC + AR(1) residuals. Top right:  A QQ plot of these residuals. Bottom left:  The sample ACF of these residuals.  Bottom right:  The sample PACF of these residuals.}
	\label{resid1}
\end{figure}

Moving to a residual analysis, Figure \ref{resid1} shows diagnostics for the TSMC marginal with an  AR(1) $\{ Z_t \}$.  The top left plot shows the raw residuals and the bottom left and right plots show sample ACFs and PACFs of these residuals.  No major issues are seen. The top right graph shows a QQ plot of these residuals for a standard normal distribution.  Some departure from normality is noted in the two tails of the plot.

\begin{figure}[h!]
	\centering
	\includegraphics[width=14cm]{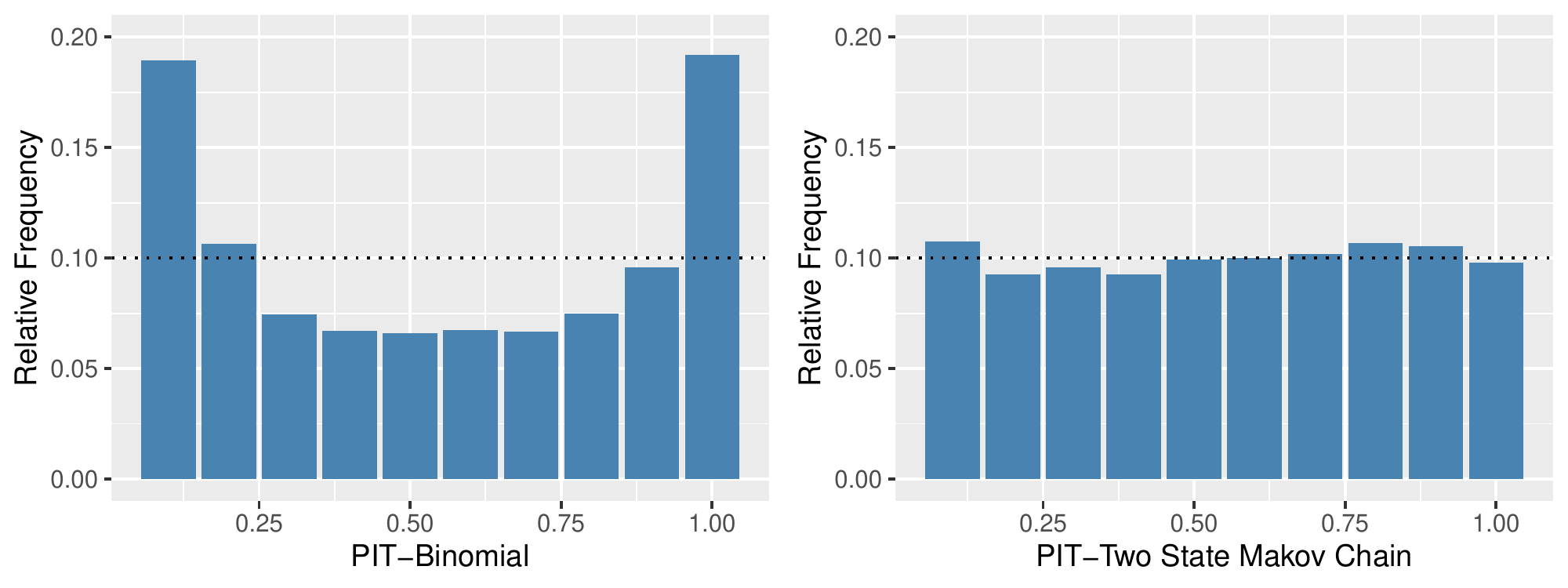}
\caption{Left: a binomial marginal PIT histogram.   Right: a TSMC marginal PIT histogram. The binomial marginal does not fit as well as the TSMC marginal.}
	\label{resid2}
\end{figure}

Finally, Figure \ref{resid2} shows PIT histograms of the residuals for the binomial and TSMC fits with AR(1) $\{ Z_t \}$.  There are obvious departures from uniformity for the binomial marginal --- this marginal distribution does not seem to describe the data well.   The histogram for the two-state Markov chain marginal is roughly uniform; hence, we have a good fitting model and the slight lack of normality in $\{ Z_t \}$ in Figure \ref{resid1} does not appear overly problematic. 

\section{Concluding Comments}

The above paper constructs a very general model for seasonal count time series through a latent Gaussian process transformation.  Any marginal distribution can be achieved and the correlations are as flexible as possible and can be negative.  Estimation of the model parameters through likelihood techniques can be conducted by particle filtering techniques.  The methods were shown to to work well on simulated data and capably handled a periodic count sequence supported on $\{ 0, 1, 2, 3, 4, 5, 6, 7 \}$.  There, we found that the latent Gaussian process did not need to be periodic, but the marginal distribution of the data contained periodicities.  The fitted model was very parsimonious, containing only seven parameters.

Extensions of the above techniques to handle covariates merit exploration. For this, we suggest allowing $\boldsymbol{\theta}$ to depend on the covariates as in \cite{jia2021count}. Trying to modify the latent Gaussian process to handle covariates generally causes trouble.   Multivariate versions of the methods are also worth studying.



\section{Appendix}
{\bf Proof of Proposition 3.1:} We follow similar reasoning to \cite{pipiras2017long} and \cite{jia2021count}.  We begin with a generalization of the Price Theorem (Theorem 5.8.5 in \cite{pipiras2017long}), stated as follows and easily proven. Let $G_{\nu_1}$ and $G_{\nu_2}$ be two continuous differentiable functions.  Then their link function has a derivative with form
\begin{equation}
\label{e:price}
L'(u) = \frac{1}{\sqrt{\mbox{Var}(X_1) \mbox{Var}(X_2)}} E [ G_{\nu_1}'(Z_1) G_{\nu_2}'(Z_2) ] \Big|_{\mbox{\scriptsize Corr}(Z_1,Z_2)=u}.
\end{equation}
Here, $Z_1$ and $Z_2$ are a correlated Gaussian pair, each component standardized, and with $\mbox{Corr}(Z_1,Z_2)=u$. 

In our application, $G_{\nu_1}$ and $G_{\nu_2}$ are non-negative and non-decreasing since they are cumulative distribution functions.  But because our data are counts, $G_{\nu_1}$ and $G_{\nu_2}$ are step functions and not necessarily differentiable on the integers.  To remedy this, we approximate $G_{\nu_1}$ and $G_{\nu_2}$ by differentiable functions and take limits in the approximation. 

To do this, let $U \stackrel{\cal D}{=}{\cal N}(0,1)$.  For any $\epsilon >0$ and $\ell \in \{ 1, 2\}$,
\begin{eqnarray}
	G_{\epsilon,\nu_\ell}(x) & := & E [ G_{\nu_\ell}(x + \epsilon U)] = \int_{- \infty}^\infty G_{\nu_\ell}(z)
	\frac{e^{-\frac{(x-z)^2}{2\epsilon^2}}}{\sqrt{2 \pi} \epsilon} dz \nonumber \\
	& = & \sum_{j=0}^\infty j \int_{\Phi^{-1}(C_{j-1}(\nu_\ell))}^{\Phi^{-1}(C_j(\nu_\ell))}
	\frac{e^{-\frac{(x-z)^2}{2\epsilon^2}}}{\sqrt{2 \pi} \epsilon} dz \nonumber \\
	& = & \sum_{j=0}^\infty j \int_{\Phi^{-1}(C_{j-1}(\nu_\ell))-x}^{\Phi^{-1}(C_j(\nu_\ell))-x}
	\frac{e^{-\frac{w^2}{2\epsilon^2}}}{\sqrt{2 \pi}\epsilon} dw.
	\label{e:G-epsilon}
\end{eqnarray}
The ``kernel'' 
\begin{equation}
\label{addon1}
\frac{e^{-\frac{(x-z)^2}{2\epsilon^2}}}
{\sqrt{2\pi}\epsilon}
\end{equation}
acts like Dirac's delta function $\delta_{\{ x \}}(z)$ at $z=x$ as $\epsilon \downarrow 0$. Note that $G_{\epsilon,\nu_\ell}(x)$ is non-decreasing and differentiable with first derivative
\begin{equation}
	\label{e:G-epsilon-derivative}
	G_{\epsilon,\nu_\ell}'(x) = \frac{1}{\sqrt{2\pi}\epsilon} \sum_{j=0}^\infty j \Big[ e^{-\frac{(\Phi^{-1}(C_{j-1}(\nu_{\ell}))-x)^2}{2\epsilon^2}} -  e^{-\frac{(\Phi^{-1}(C_{j}(\nu_{\ell}))-x)^2}{2\epsilon^2}} \Big] = \frac{1}{\sqrt{2\pi}\epsilon} \sum_{j=0}^\infty e^{-\frac{(\Phi^{-1}(C_{j}(\nu_{\ell}))-x)^2}{2\epsilon^2}},
\end{equation}
and define $X_{\ell}^{(\epsilon)} = G_{\epsilon,\nu_\ell}(Z_{\ell})$ for $\ell \in \{ 1, 2 \} $. Equation (\ref{e:link-derivative-again}) gives
\begin{eqnarray}
L_\epsilon'(u) & = & \frac{1}{\sqrt{\mbox{Var}(X_1^{(\epsilon)})\mbox{Var}(X_2^{(\epsilon)})}}E [G_{\epsilon,\nu_1}'(Z_1) G_{\epsilon,\nu_2}'(Z_2) ] \Big|_{\mbox{\scriptsize Corr}(Z_1,Z_2)=u} \nonumber \\
	& = & \frac{1}{\sqrt{\mbox{Var}(X_1^{(\epsilon)})\mbox{Var}(X_2^{(\epsilon)})}}
	\int_{-\infty}^\infty
	\int_{-\infty}^\infty G_{\epsilon,\nu_1}'(Z_1) G_{\epsilon,\nu_2}'(Z_2)\frac{1}{2\pi\sqrt{1-u^2}} e^{-\frac{1}{2(1-u^2)}\big(z_1^2 + z_2^2 - 2 u z_1 z_2\big)} dz_1dz_2 \nonumber \\
	& = &  \frac{1}{\sqrt{\mbox{Var}(X_1^{(\epsilon)})\mbox{Var}(X_2^{(\epsilon)})}} \sum_{j_1=0}^\infty \sum_{j_2=0}^\infty \int_{-\infty}^\infty  \int_{- \infty}^\infty
	\frac{1}{\sqrt{2\pi}\epsilon} e^{-\frac{(\Phi^{-1}(C_{j_1}(\nu_1))-z_1)^2}{2\epsilon^2}}
	\frac{1}{\sqrt{2\pi}\epsilon} e^{-\frac{(\Phi^{-1}(C_{j_2}(\nu_2))-z_2)^2}{2\epsilon^2}} \times \nonumber \\
	& & \quad \quad  \ \frac{1}{2\pi\sqrt{1-u^2}} e^{-\frac{1}{2(1-u^2)}\big(z_1^2 + z_2^2 - 2 u z_1 z_2\big)} dz_1dz_2.
\end{eqnarray}
Noting again that the quantity in (\ref{addon1}) acts like a Dirac's delta function $\delta_{\{x\}}(z)$, the limit as $\epsilon \downarrow 0$ should be
\begin{equation}
L'(u) = \frac{1}{\sqrt{\mbox{Var}(X_1)\mbox{Var}(X_2)}}  \sum_{j_1=0}^\infty \sum_{j_2=0}^\infty \frac{1}{2\pi \sqrt{1-u^2}}
	e^{-\frac{1}{2(1-u^2)}\big( \Phi^{-1}(C_{j_1}(\nu_1))^2 + \Phi^{-1}(C_{j_2}(\nu_2))^2 - 2 u \Phi^{-1}(C_{j_1}(\nu_1)) \Phi^{-1}(C_{j_2}(\nu_2))\big)},
\end{equation}
which is always non-negative. The existence and form of $L^\prime(u)$ stems from the fact that we are differentiating a power series with absolutely convergent coefficients inside its radius of convergence.  That $\sum_{k=0}^\infty |\ell_k| < \infty$ follows from (\ref{eq:link_coefficient}), the Cauchy-Schwarz inequality, and the finiteness of $\sum_{k=0}^\infty k! g_k(\nu_1)^2$ and $\sum_{k=0}^\infty k!g_k(\nu_2)^2$.

We now show that $L_\epsilon'(u)$ converges to $L'(u)$. For this, we first need an expression for the Hermite coefficients of $G_{\epsilon,\nu_\ell}(\cdot)$, denoted by $g_{\epsilon,k}(\nu_\ell)$ for $\ell \in \{ 1, 2 \}$.   These will be compared to the Hermite coefficients $ g_{k}(\nu_\ell) $ of $G_{\nu_\ell}$. 

Taylor expanding the Hermite polynomial $H_k(x+y) = \sum_{d=0}^k {k \choose d} y^{k-d} H_d(x)$ implies
\begin{eqnarray*}
	G_{\epsilon,\nu_\ell}(x) & = &  E [G_{\nu_\ell}(x+\epsilon U)] =
	E \left[ \sum_{k=0}^\infty g_{k}(\nu_\ell) H_k(x+\epsilon U) \right] \\
	& = &  E \left[ \sum_{k=0}^\infty g_{k}(\nu_\ell) \sum_{d=0}^k {k \choose d} (\epsilon U)^{k-d} H_d(x) \right] \\
	& = & \sum_{d=0}^\infty H_d(x) \sum_{k=d}^\infty g_{k}(\nu_\ell) \epsilon^{k-d} {k \choose d}  E[U^{k-d}].
\end{eqnarray*}
After changing summation indices and using that $E[U^p] = 0$ if $p$ is odd, and equal to $(p-1)!!$ if $p$ is even, where $k!!=1 \times 3 \times \cdots \times k$ when $k$ is odd, we get
\begin{equation}
\label{e:gk-epsilon}
g_{\epsilon,k}(\nu_\ell) = g_{k}(\nu_\ell) + \sum_{q=1}^\infty g_{k+2q}(\nu_\ell) \epsilon^{2q} {k+2q \choose k} (2q-1)!! =  g_{k}(\nu_\ell) + \sum_{q=1}^\infty g_{k+2q}(\nu_\ell) \epsilon^{2q}\frac{(k+2q)!}{k!2^qq!}.
\end{equation}
Then
\begin{eqnarray}
\label{e:gk1gk2-epsilon}
    \nonumber
	|g_{k}(\nu_1)g_{k}(\nu_2) - g_{\epsilon,k}(\nu_1)g_{\epsilon,k}(\nu_2)| &\leq & |g_{k}(\nu_1)| \sum_{q=1}^\infty |g_{k+2q}(\nu_1)| \epsilon^{2q}\frac{(k+2q)!}{k!2^qq!} + |g_{k}(\nu_2)| \sum_{q=1}^\infty |g_{k+2q}(\nu_2)| \epsilon^{2q}\frac{(k+2q)!}{k!2^qq!} \\
	&&+ \Big(\sum_{q=1}^\infty |g_{k+2q}(\nu_1)| \epsilon^{2q}\frac{(k+2q)!}{k!2^qq!}\Big) \Big(\sum_{q=1}^\infty |g_{k+2q}(\nu_2)| \epsilon^{2q}\frac{(k+2q)!}{k!2^qq!}\Big).
\end{eqnarray}
Use the Cauchy-Schwarz inequality to obtain the bounds
\[
\sum_{q=1}^\infty |g_{k+2q}(\nu_\ell)| \epsilon^{2q}\frac{(k+2q)!}{k!2^qq!} \leq
\left( \sum_{q=1}^\infty g_{k+2q}^2(\nu_\ell) (k+2q)! \right)^{1/2}
\left( \sum_{q=1}^\infty \epsilon^{4q} \frac{(k+2q)!}{(k!)^2(2^qq!)^2} \right)^{1/2}
\]
\[
\leq \frac{M_{k,\ell}}{(k!)^{1/2}}
\left(  \sum_{q=1}^\infty \epsilon^{4q} \frac{(k+2q)!}{k!(2q)!} \right)^{1/2}\quad\quad \forall\ell\in\{1,2\} ,
\]
where $M_{k,\ell}$ is some finite constant that converges to zero as $k \rightarrow \infty$. Since $\mbox{\rm Var}(X_\ell) = \sum_{k=1}^\infty k! g_{k}^2(\nu_\ell)$ is finite and $(2^qq!)^2$ is of the same order as $(2q)!$, $\sum_{q=1}^\infty g_{k+2q}^2(\nu_\ell) (k+2q)! \to 0$ as $k \rightarrow \infty$. We use the fact that $\sum_{p=0}^\infty x^{p} {k+p\choose p} = (1-x)^{-k-1}$ for $|x|<1$ to obtain a bound for $\sum_{p=1}^\infty \epsilon^{2p} {k+p \choose p}$. Then (\ref{e:gk1gk2-epsilon}) gives
\begin{equation}
	\label{e:gk-epsilon-gk-2}
	|g_{k}(\nu_1)g_{k}(\nu_1) - g_{\epsilon,k}(\nu_1)g_{\epsilon,k}(\nu_2)|  \leq  \sum_{\ell=1}^{2}\frac{ M_{k,\ell}|g_{k}(\nu_\ell)|}{(k!)^{1/2}} \left[ (1-\epsilon^2)^{-k-1} -1 \right]^{1/2} + \frac{M_{k,1}M_{k,2}}{k!} [(1-\epsilon^2)^{-k-1} -1].
\end{equation}
Now take the first derivative of the link function in (\ref{eq:link}) to obtain
\[
L'(u) = \frac{1}{\sqrt{\mbox{Var}(X_1)\mbox{Var}(X_2)}} \sum_{k=1}^\infty g_{k}(\nu_1)g_{k}(\nu_2)k!ku^{k-1},
\]
where the series converges absolutely for $u \in (-1,1)$ since the ``extra'' $k$ gets dominated by $u^{k-1}$. Similarly,
\[
L_\epsilon'(u) =  \frac{1}{\sqrt{\mbox{Var}(X_1)\mbox{Var}(X_2)}}\sum_{k=1}^\infty g_{\epsilon,k}(\nu_1)g_{\epsilon,k}(\nu_2)k!ku^{k-1}.
\]
The above expression agrees with Theorem 5.1.10 in \cite{pipiras2017long}. To show that the difference between $L'_{\epsilon}(u)$ and $L'(u)$ converges to zero as $\epsilon \downarrow 0$, use
\begin{eqnarray}
	\nonumber
	&&|L'(u) - L_\epsilon'(u)| \leq \Big| \frac{1}{\sqrt{\mbox{Var}(X_1)\mbox{Var}(X_2)}}- \frac{1}{\sqrt{\mbox{Var}(X_1^{(\epsilon)})\mbox{Var}(X_2^{(\epsilon)})}}\Big|  \sum_{k=1}^\infty g_{k}(\nu_1)g_{k}(\nu_2)k!  k|u|^{k-1} \\
	&&+\frac{1}{\sqrt{\mbox{Var}(X_1^{(\epsilon)})\mbox{Var}(X_2^{(\epsilon)})}} \sum_{k=1}^\infty |g_{k}(\nu_1)g_{k}(\nu_2) - g_{\epsilon,k}(\nu_1) g_{\epsilon,k}(\nu_2)|k! k|u|^{k-1}.
\end{eqnarray}
From (\ref{e:gk-epsilon-gk-2}), we see that $|g_{k}(\nu_1)g_{k}(\nu_2) - g_{\epsilon,k}(\nu_1)g_{\epsilon,k}(\nu_2)| \rightarrow 0$ as $\epsilon \downarrow 0$. Hence,  $\sum_{k=1}^\infty |g_{k}(\nu_1)g_{k}(\nu_2) - g_{\epsilon,k}(\nu_1) g_{\epsilon,k}(\nu_2)|k! k|u|^{k-1}$ converges to zero by the dominated convergence theorem as $\epsilon \downarrow 0$. Using (\ref{link_correlation}), we concluded that $\mbox{Var}(X_1^{(\epsilon)})\rightarrow \mbox{Var}(X_1)$ and $\mbox{Var}(X_2^{(\epsilon)}) \rightarrow \mbox{Var}(X_2)$ as $\epsilon \downarrow 0$. Therefore,
\[
\left| 
\frac{1}{\sqrt{\mbox{Var}(X_1)\mbox{Var}(X_2)}}-
\frac{1}{
\sqrt{\mbox{Var}(X_1^{(\epsilon)})
\mbox{Var}(X_2^{(\epsilon)})}
}
\right| 
\rightarrow 0
\quad \mbox{as} \quad \epsilon \downarrow 0
\]
follows by continuity of the function $x^{-1/2}$ away from $x=0$ (the limiting variances are tacitly assumed positive to avoid degeneracy). 

\bibliography{References}

\end{document}